\documentclass[pre,twocolumn,superscriptaddress]{revtex4}
\usepackage{graphicx,subfigure}
\usepackage{amsthm}
\usepackage{amsmath}
\usepackage{rotating}
\usepackage{color,colordvi}
\newcommand{\Epsilon}{\mathcal{E}} 
\newcounter{reaction} 
\renewcommand{\thereaction}{R\arabic{reaction}}

\newcommand{\tabreaction}{\protect\refstepcounter{reaction}(\thereaction)}

\begin{document}
\bibliographystyle{revtex4}


\title{On the electron energy distribution function in the high power impulse magnetron sputtering discharge}



\author{Martin Rudolph}
\email[]{martin.rudolph@iom-leipzig.de}

\affiliation{Leibniz Institute of Surface Engineering (IOM), Permoserstra{\ss}e 15, 04318 Leipzig, Germany}

\author{A. Revel}

\affiliation{Laboratoire de Physique des Gaz et Plasmas - LPGP, UMR 8578 CNRS, 
Universit{\'e} Paris--Saclay,  91405 Orsay Cedex, France}

\author{D. Lundin}

\affiliation{Laboratoire de Physique des Gaz et Plasmas - LPGP, UMR 8578 CNRS, 
Universit{\'e} Paris--Saclay,  91405 Orsay Cedex, France}
\affiliation{Plasma and Coatings Physics Division, IFM-Materials Physics,
 Link{\"o}ping University, SE-581 83,  Link{\"o}ping, Sweden}

\author{H. Hajihoseini}

\affiliation{Science Institute, University of Iceland, Dunhaga 3, IS-107 Reykjavik, Iceland}
\affiliation{Industrial Focus Group XUV Optics, MESA+ Institute for Nanotechnology, University of Twente, Drienerlolaan 5,
7522 NB Enschede, The Netherlands}

\author{N. Brenning}

\affiliation{Plasma and Coatings Physics Division, IFM-Materials Physics,
 Link{\"o}ping University, SE-581 83,  Link{\"o}ping, Sweden}

\affiliation{Department of Space and Plasma Physics, School of Electrical Engineering and Computer Science,  KTH  Royal Institute of Technology, SE-100 44, Stockholm, Sweden}

\author{M. A. Raadu}

\affiliation{Department of Space and Plasma Physics, School of Electrical Engineering and Computer Science,  KTH  Royal Institute of Technology, SE-100 44, Stockholm, Sweden}

\author{A. Anders}

\affiliation{Leibniz Institute of Surface Engineering (IOM), Permoserstra{\ss}e 15, 04318 Leipzig, Germany}

\affiliation{Felix Bloch Institute of Solid State Physics, Leipzig University, Linn{\'e}stra{\ss}e 5, 04103 Leipzig, Germany}

\author{T. M. Minea}

\affiliation{Laboratoire de Physique des Gaz et Plasmas - LPGP, UMR 8578 CNRS, 
Universit{\'e} Paris--Saclay,  91405 Orsay Cedex, France}

\author{J. T. Gudmundsson}
\email[]{tumi@hi.is}


\affiliation{Science Institute, University of Iceland, Dunhaga 3, IS-107 Reykjavik, Iceland}

\affiliation{Department of Space and Plasma Physics, School of Electrical Engineering and Computer Science,  KTH  Royal Institute of Technology, SE-100 44, Stockholm, Sweden}

\date{\today}

\begin{abstract}
We apply the Ionization Region Model (IRM) and the Orsay  Boltzmann equation  for  ELectrons  coupled  with  Ionization and  eXcited  states  kinetics
(OBELIX) model to study the electron kinetics of a high power impulse magnetron sputtering (HiPIMS) discharge.  In the IRM the bulk (cold)  electrons are assumed to exhibit a Maxwellian energy distribution and the secondary (hot)  electrons, emitted from the target surface upon ion bombardment, are treated as a high energy tail, 
while in the  OBELIX the electron energy distribution  is calculated self-consistently using an isotropic Boltzmann equation.  The two models are merged in the sense that the output from the IRM is used as an input for OBELIX. The temporal evolutions of the particle densities are found to agree very well between the two models.  
Furthermore, a  very good agreement is demonstrated  between the bi-Maxwellian electron energy distribution assumed by the IRM and the  electron energy distribution calculated by the OBELIX model.  It can therefore be concluded that assuming a bi-Maxwellian electron energy distribution, constituting a cold bulk electron group  and a hot secondary electron group, is a good approximation for modeling the HiPIMS discharge. 
\end{abstract}

\maketitle


\section{Introduction}

When a magnetron sputtering discharge \citep{gudmundsson20:113001} is driven by  high-power  unipolar voltage pulses applied to the cathode target at  low repetition frequency and low   duty  cycle \citep{gudmundsson12:030801,hubicka20:49},  while keeping  the  average  power  about  two  orders  of  magnitude lower than the pulse peak power \citep{anders11:S1}, 
it is referred to as a high-power impulse magnetron sputtering (HiPIMS) discharge \citep{gudmundsson12:030801}.  The HiPIMS discharge is based on  essentially the same apparatus as the dc magnetron sputtering (dcMS) discharge \citep{gudmundsson20:113001} except for the addition of a pulser unit \citep{hubicka20:49}.  HiPIMS  is an ionized physical vapor deposition (IPVD) technique that has been demonstrated to deposit thin films with improved properties compared to thin films deposited by dcMS.  The improved thin film properties have been  related to 
a very high ionization fraction of the sputtered species \citep{lundin12:780,greczynski19:060801,greczynski20:180901}, which is a consequence of a high electron density \citep{gudmundsson20:113001,gudmundsson12:030801}. 

It is well established that low-energy ion bombardment has a beneficial influence on the microstructure of the growing film, as it  enhances the adatom surface mobility \citep{petrov93:36,petrov03:S117}.  However, more important is the bombardment of the growing film by ions of the film-forming material as they can eliminate film porosity when depositing at low substrate temperatures since they are primarily incorporated at lattice sites \citep{greczynski19:060801,greczynski20:180901}. 
  Furthermore, the high ionization fraction of the film-forming material allows for better control over the thin film growth, as it is possible to control the energy and direction of the sputtered species by applying a substrate bias \citep{hubicka20:49}  and therefore  tuning the thin film material properties, such as  hardness, surface roughness, crystallinity, preferred orientation, refractive index,  and residual stress \citep{lundin12:780,alami05:278, greczynski19:060801,greczynski20:180901,sarakinos20:333,cemin17:120,hajihoseini18:126}.  This is a significant advantage over deposition by dcMS  where the film-forming material consists mainly of neutral species and the ions reaching the substrate are mainly ions of the noble working gas \citep{gudmundsson20:113001}.

In the case of low pressure, low temperature, partially ionized plasma discharges, such as the HiPIMS discharge, the electrons are generally not in thermal equilibrium with the heavy species, and they are also often  not in thermal equilibrium among themselves.  In fact, it has been observed experimentally that the bulk ($<$ 10 eV) electron energy distribution function (EEDF) is  either Maxwellian-like or bi-Maxwellian-like, depending on the working gas pressure and spatial
location within the dcMS discharge \citep{sheridan91:688,seo04:409} and the HiPIMS discharge  \citep{gudmundsson02:249,pajdarova09:025008}.  
Furthermore, secondary electrons, created by ion bombardment of the cathode target, are even more energetic ($> 100$ eV), since they are accelerated across the cathode sheath.
It is of significant importance to determine, be able to predict, and even control, the EEDF in the HiPIMS discharge, as the electrons dictate the ionization processes for both the working gas and the sputtered species and therefore determine the discharge properties \citep{brenning17:125003}. 
 
There have been a few attempts to model the HiPIMS discharge using various approaches to treat the electron population \citep{minea20:159}.  The simplest approach is based on following the sputtered and working gas species within the discharge, referred to as  the phenomenological material pathways model, set forth by \citet{christie05:330}, and later developed further by   Vl\v{c}ek et al.~\citep{vlcek07:42,vlcek10:065010}.  Using this model, the fraction of the target material ions that  return to the target and therefore do not contribute to the deposited film, can be evaluated.  Also, it provides relations between the applied target power density, the  deposition rate, and the flux of ions bombarding the growing film.   This modeling approach has turned out to be  very important for understanding the low deposition rate and various other issues related to the operation of the HiPIMS discharge.
  However, in this approach the electrons are not followed and no information is gained on the properties of the electrons.  

The ionization region model (IRM) is a volume-averaged, time-dependent, plasma chemical model,  that describes the ionization region (IR) of the HiPIMS discharge \citep{raadu11:065007,huo17:354003}.  Using this model, two main electron power absorption mechanisms that drive  the magnetron sputtering discharge, have been identified; sheath energization of secondary electrons and Ohmic heating within the ionization region \citep{huo13:045005,brenning16:065024}.  
The first power absorption mechanisms is due to the acceleration of secondary electrons across the cathode sheath that forms between the cathode target surface and the IR.  In dcMS operation, where the discharge current at the cathode target surface is mostly carried by Ar$^+$ ions  sheath energization can exhibit significant contribution for target materials with  high secondary electron emission coefficients  $\gamma_{\rm see}$ \citep{depla09:2825,brenning16:065024}.  The sheath acceleration basically injects high energy  secondary electrons, emitted from the target due to ion bombardment, into the IR \citep{thornton78:171}, just like in the dc glow discharge that is maintained by the emission of secondary electrons \citep{gudmundsson17:123001}. However, the more important electron power absorption mechanism in HiPIMS operation is usually Ohmic heating of the electrons due to a potential drop that develops across the ionization region (typically few tens of volts) \citep{huo13:045005,huo17:354003,brenning16:065024,panjan17:063302,rudolph21:033303}.   In HiPIMS operation, singly charged ions of the sputtered material typically exhibit $\gamma_{\rm see} \approx 0$ \citep{anders08:201501}, while bombardment by Ar$^+$ ions and doubly charged metal ions can contribute to emission of secondary electrons.  For HiPIMS operation significant portion of the ions bombarding the cathode target are ions of the sputtered species \citep{brenning17:125003,huo17:354003}.
  The cathode potentials in a HiPIMS discharge can go up to several hundred  volts or higher. Consequently, the energy of the secondary electrons is significantly higher than the typical peak of the thermalized component of the bulk electrons. These two very different energy ranges are reflected, in the ionization region model, as two electron energy distributions, describing the hot and cold electron groups.  For simplicity, in the IRM, the bulk (cold)  electrons are assumed to exhibit a Maxwellian  energy distribution, while the secondary electrons  appear as a high-energy tail 
\citep{huo17:354003}.  Here we explore  how accurate this approximation is.

There are also attempts to study the EEDF in HiPIMS in more details using model approaches. 
\citet{gallian15:023305} presented analytic calculations of the energy distribution function for the energetic  electrons in the magnetron sputtering discharge by describing them as an initially monoenergetic beam that slows down through Coulomb
collisions with  Maxwellian distributed  bulk (cold) electrons, and through inelastic collisions with
neutrals.   They  provide an analytic solution for the energetic electron energy distribution that they claim can be used to calculate correction terms for fluid descriptions of the electron species. Using particle-in-cell Monte Carlo collision (PIC/MCC) simulations the EEDF can be calculated self-consistently along with spatial and temporal variation of the various plasma parameters. However, due to the timescales involved and the high electron densities, it remains challenging to apply PIC/MCC simulations to the HiPIMS discharge \citep{minea20:159}, but a few attempts have been successful, including both 2D  \citep{revel18:105009} and pseudo-3D \citep{revel16:100701} PIC/MCC simulation.
Using a 2D  PIC/MCC simulation \citet{revel18:105009} observe a EEDF that is composed of at least two Maxwellian-like  distributions during the pulse.
 In fact, it is clear from  these studies and the experimental findings discussed earlier that a kinetic approach should be pursued to determine the EEDF when  modeling such a discharge.

These issues are addressed in the  Orsay  Boltzmann equation  for  ELectrons  coupled  with  Ionization  and  eXcited  states  kinetics (OBELIX) model, which is a volume-averaged collisional-radiative model  where the EEDF in a HiPIMS discharge is calculated self-consistently solving an isotropic Boltzmann equation \citep{guimaraes93:127}.  It was originally developed as a collisional radiative model of an argon glow discharge \citep{bretagne81:1225,bretagne82:2205,bretagne86:761} and later modified to model a dcMS discharge in argon with a molybdenum target \citep{guimaraes91:133,guimaraes93:127}, an Ar/O$_2$ mixture with a chromium target \citep{trennepohl96:607}, and finally a HiPIMS discharge in argon with a  copper target \citep{bretagne15:P1.17}. For this work, we have thoroughly revised OBELIX and added, in particular, a detailed description of the argon energy levels, including excitation and de-excitation reactions as well as radiative transitions between the energy levels and ionization from each level. 

Here, we merge OBELIX with the IRM to make use of the self-consistent calculation of the EEDF.  This allows us to compare the results from IRM with the results from OBELIX.   The IRM is described in Section \ref{irmsection} and the OBELIX model is described in Section \ref{obelixsection}.  The results from the two models, when applied to a HiPIMS discharge in argon with Ti target, are demonstrated and compared in Section \ref{resultsanddiscussion}. We compare the temporal evolution of the various species and the energy cost of ionization, as well as the EEDF, resulting from these two approaches. The main goal of this work is to assess if the loss of accuracy by using the IRM is acceptable, given the gain in computational time. The findings are summarized in Section \ref{conclusion}.  

\section{The ionization region model}
\label{irmsection}

\subsection{General description}
\label{IRMdescript}

The ionization region model (IRM) is a volume-averaged time-dependent plasma chemical model of the
 ionization region (IR) within the HiPIMS discharge \citep{raadu11:065007,huo17:354003}.  
The IR constitutes the  brightly glowing plasma torus located next to the target race track, that appears due to magnetic confinement of the electrons.
In the model, the IR is described as an annular cylinder  with outer radii $r_{\rm c2}$, and inner radii $r_{\rm c1}$, marking the race track region, and length $ L = z_2 - z_1$, extending from $z_1$ to $z_2$ axially away from the target. Here $z_1$ is the cathode sheath thickness and $z_2$ marks the extension of the IR (see e.g.~Figure 1 in \citet{raadu11:065007}).  The model assumes only volume-averaged values of the electron, ion and neutral densities and the electron temperature over the IR volume.   The temporal development of the plasma parameters is defined by a
set of ordinary differential equations and the electron density is determined assuming quasi-neutrality of the
plasma discharge \citep{huo17:354003}.

The IRM  has  been applied to study gas rarefaction and refill processes \citep{raadu11:065007,huo12:045004},
the reduction in deposition rate \citep{brenning12:025005}, the ion composition at the target surface \citep{huo17:354003},
 and the electron heating mechanism \citep{huo13:045005}
in an argon HiPIMS  discharge with an Al target. For a Ti target, the ion composition at the
target surface \citep{huo17:354003}, the  temporal behavior of the argon metastable states  \citep{stancu15:045011} and their role and importance in the ionization processes  \citep{gudmundsson15:113508},  as well as  the effect of shortening the pulse length on the ionized flux fraction and deposition rate \citep{rudolph20:05LT01}, have been studied.   

\subsection{Fitting procedure}

The IRM is semi-empirical and is constrained by experimental data. First the model needs to be adapted to a physical discharge (the geometry, the working gas, the working gas pressure, working gas species,
 target material and target species, sputter yields, and a reaction set for these species)  \citep{huo17:354003}.  Then two or three parameters are adjusted such that the model reproduces the measured discharge current and cathode voltage waveforms,  $I_{\rm D}(t)$ and $V_{\rm D}(t)$, respectively \citep{huo17:354003}.  The fitting parameters are the potential  drop across the ionization region $V_{\rm IR}$ and the ion back attraction probability of the sputtered species $\beta_{\rm t}$.  
    These  parameters are adjusted so that the current to the cathode target calculated in the model best reproduces an experimentally determined discharge current. 
Recently, it has been demonstrated how the measured ionized flux fraction   of the sputtered species can be used to further constrain the model \citep{butler18:105005}.
Together, the two constraints provided by the measured  discharge current waveform and the ionized flux fraction lock the  values for the discharge voltage that drops over the IR, $V_{\rm IR}$, and the back-attraction probability of the sputtered species $\beta_{\rm t}$ \citep{butler18:105005}.  
The voltage that drops across the IR, $V_{\rm IR}$,  accounts for the Ohmic power transfer to the cold electrons.
The discharge voltage $V_{\rm D}$ is therefore split between the cathode sheath and the IR such that  $V_{\rm D} = V_{\rm SH} + V_{\rm IR}$, where $V_{\rm SH}$ is the voltage drop across the cathode sheath.
The other fitting parameter $\beta_{\rm t}$, accounts for the back-attraction probability of the ions of the sputtered species to the cathode target.  The incorporation of the back-attraction probability into the IRM was inspired by the materials pathways model \citep{brenning12:025005}. 
Sometimes  (typically not for metallic targets),  a third adjustable parameter is needed, the probability of return and recapture of 
secondary electrons emitted from the target,  caused by the magnetic field, and denoted by $r$  \citep{gudmundsson16:065004}. 
 Note that the  magnetic field was taken into account in the IRM, albeit not directly but via the input of the measured discharge current  and voltage which depend on the magnetic field and therefore the modeled discharge properties such as species densities and fluxes in and out of the ionization region.    Comparison of the internal discharge parameters, the target species ionization probability $\alpha_{\rm t}$ and the target ion back-attraction probability $\beta_{\rm t}$ determined by the IRM and an analytical model using measured deposition rates and ionized flux fractions show excellent agreement \citep{rudolph21:033303}. 

\subsection{The reaction sets for IRM}
\label{reactionsetsIRM}

Here, for this current study, we take argon as the working gas and titanium as the cathode target material and use the IRM 
to determine the time-varying density of the modeled species (Ti, Ti$^+$, Ti$^{2+}$, Ar, Ar$^+$, metastable species Ar$^{\rm m}$, and hot and cold electrons) within the ionization region (IR) as well as the time-varying fluxes of these species in and out of the ionization region, e.g.~from and to the cathode target race track, and from and to the diffusion region (DR).  

The IRM is based  on rate coefficients that are calculated using an assumed EEDF that constitutes  cold and hot electron groups each having a Maxwellian distribution. 
The Maxwellian population of the cold electrons constitutes the majority of the electrons and therefore dictates the electron density and sets the effective electron temperature in the few eV range.  To describe this, the IRM uses two sets of rate coefficients, one for a cold and another for a  hot electron  group.   
The rate coefficients for the cold electrons were determined assuming a Maxwellian EEDF and fit in the
electron temperature range 1 -- 7 eV, and for the  hot electrons the EEDF is also assumed to be Maxwellian and fit in the electron temperature range 200 -- 1000 eV. These rate coefficients are applied to the entire range in electron energy.
The reaction set and the rate coefficients here included in the IRM  are mostly the same as used in our earlier work on HiPIMS discharges with a titanium cathode target \citep{stancu15:045011,gudmundsson16:065004,huo17:354003}.   However, a few modifications of the rate coefficients have been made, mainly regarding the 
metastable argon atoms.  The electron impact excitation rate coefficients into the 4s levels of the argon atom are calculated based on the 
cross sections from the IST-Lisbon collection on LXCat \citep{ISTLisbon}  and discussed by \citet{alves14:012007} and
\citet{yanguasgil05:1588}.  The cross sections  originate from the work of \citet{khakoo04:247}. 
The levels 
Ar(4s'[1/2]$_0$),  and 
Ar(4s[3/2]$_2$) (in Racah's notation) are metastable and each level is now included in the IRM as separate species. 
 The electron impact ionization rate coefficient from the metastable levels is  calculated from the  cross section measured by \citet{dixon73:405} (see also  \citet{freund87:329}). The cross sections used to calculate the rate coefficients in the IRM are compared to the cross sections used in the OBELIX in Appendix \ref{crosssectionappendix}.
 The rate coefficients for electron impact de-excitation of the  metastable levels are calculated by applying the principle of detailed balancing \citep[Section 8.5]{lieberman05}.   All the reactions and rate coefficients included in the IRM for this current study  are listed in Table \ref{ratecoeff}. 
\begin{table*}
\caption{The reactions and  rate coefficients used in the IRM for a discharge with argon as the working gas and titanium cathode target including both hot and cold electrons.  The rate coefficients are calculated assuming a Maxwellian electron energy distribution function and fit in the range $T_{\rm e} = 1 - 7$ eV for cold electrons and 200 -- 1000 eV for hot electrons. \label{ratecoeff} 
}
{\tiny
\begin{center} 
\begin{tabular}{ llllll } 
\hline \hline
  & Reaction       &  Threshold  \ \ \  &  Rate coefficient  &  electrons &  Reference  \\
  &              &  [eV]       &  [m$^3$/s]                 &       &  \\
\hline
\\
\tabreaction \label{Ra:1} & e + Ar(3p$^6$)~$\longrightarrow$~ Ar$^+$ + e + e & 15.76 & $2.34 \times 10^{-14}$ ${T_e}^{0.59}$ $\mathrm{e}^{-17.44/\mathrm{T_e}}$  &  cold & \citep{gudmundsson07:399} \\
                &  & & $8 \times 10^{-14}$ ${T_e}^{0.16}$ $\mathrm{e}^{-27.53/{T_e}}$  & hot & \\               
\tabreaction \label{Ra:2} &  e + Ar(3p$^6$) $\longrightarrow$ Ar(4s[3/2]$_2$) + e  & 11.548  &  $1.617 \times 10^{-14} T_{\rm e}^{-0.8238} \exp(-14.1256/T_{\rm e})$  &cold    & \citep{ISTLisbon}  \\ 
  &  & & $1.1397  \times 10^{-22}  T_{\rm e}^2  - 1.8975 \times 10^{-19}   T_{\rm e}  +  8.7910  \times 10^{-17}$ & hot & \\
\tabreaction \label{Ra:4} &  e + Ar(3p$^6$)  $\longrightarrow$ Ar(4s'[1/2]$_0$) + e  & 11.723  & $2.86 \times 10^{-15} T_{\rm e}^{-0.8572} \exp(-14.6219/T_{\rm e})$  & cold & \citep{ISTLisbon}  \\
& & & $1.8045 \times 10^{-23}  T_{\rm e}^2   - 2.9825 \times 10^{-20}   T_{\rm e} +   1.357 \times 10^{-17}$ & hot &  \\ 
\tabreaction \label{Ra:6} & e + Ar(4s[3/2]$_2$)  $\longrightarrow$  Ar(3p$^6$) + e  &   &  $3.23  \times 10^{-15} T_{\rm e}^{-0.8238} \exp(-2.578/T_{\rm e})$  & cold    & Detailed \\ 
& & & $(1.1397  \times 10^{-22}  T_{\rm e}^2  - 1.8975 \times 10^{-19}   T_{\rm e}  +  8.7910  \times 10^{-17})/5$ & hot & balancing \\ 
\tabreaction \label{Ra:8} &  e +  Ar(4s'[1/2]$_0$)  $\longrightarrow$   Ar(3p$^6$) + e  &   & $2.86 \times 10^{-15} T_{\rm e}^{-0.8572} \exp(-2.8989/T_{\rm e})$ & cold & Detailed   \\ 
& & & $1.8045 \times 10^{-23}  T_{\rm e}^2   - 2.9825 \times 10^{-20}   T_{\rm e} +   1.357 \times 10^{-17}$  & hot & balancing \\ 
\tabreaction \label{Ra:11} &  e + Ar(4s'[1/2]$_0$) $\longrightarrow$ Ar$^+$ + 2e & 4.21  & $1.14356 \times 10^{-13} T_{\rm e}^{0.2548} \exp(-4.4005/T_{\rm e})$  & cold &  \citep{dixon73:405,freund87:329}  \\ 
& & & $1.5213 \times 10^{-19}   T_{\rm e}^2  - 2.9599 \times 10^{-16}  T_{\rm e} +  1.8155 \times 10^{-13}$ & hot &  \\ 
\tabreaction \label{Ra:12} &  e + Ar(4s[3/2]$_2$)  $\longrightarrow$ Ar$^+$ + 2e & 4.21  & $1.14356 \times 10^{-13} T_{\rm e}^{0.2548} \exp(-4.4005/T_{\rm e})$  &cold  &  \citep{dixon73:405,freund87:329}  \\ 
& & & $1.5213 \times 10^{-19}   T_{\rm e}^2  - 2.9599 \times 10^{-16}  T_{\rm e} +  1.8155 \times 10^{-13}$ & hot &  \\ 
\tabreaction \label{Ra:13} &  e + Ti  ~$\rightarrow$~ Ti$^+$ + e  & 6.837 &  $2.8278\times 10^{-13} {T_{\rm e}}^{-0.0579}$ $\mathrm{e}^{-8.7163/{T_{\rm e}}}$  & cold & \citep{deutsch08:58,bartlett04:235} \\    
   & & &  $1.1757\times 10^{-12} {T_{\rm e}}^{-0.3039}$ $\mathrm{e}^{-21.1107/{T_{\rm e}}}$ & hot &  \\ 
\tabreaction \label{Ra:14} &   e + Ti$^+$  ~$\rightarrow$~ Ti$^{2+}$ + e  & 13.594 &  $1.8556\times 10^{-14} {T_{\rm e}}^{0.4598}$ $\mathrm{e}^{-12.9927/\mathrm{T_{\rm e}}}$  & cold & \citep{diserens88:2129} \\
& & & $8.1858\times 10^{-12} {T_{\rm e}}^{-0.669}$ $\mathrm{e}^{-200.93/{T_{\rm e}}}$ & hot &  \\ 
\tabreaction \label{Ra:15} &  Ar$^{\rm +}$ + Ti   ~$\rightarrow$~  Ar + Ti$^+$ & & $1 \times 10^{-15}$ & &  \\     
\tabreaction \label{Ra:16} &  Ar(4s'[1/2]$_0$) + Ti  ~$\rightarrow$~  Ar + Ti$^+$ + e  & & $3.2 \times 10^{-15}$  & &  \citep{riseberg73:1962,stancu15:045011} \\   
\tabreaction \label{Ra:17} &  Ar(4s[3/2]$_2$) + Ti  ~$\rightarrow$~  Ar + Ti$^+$ + e  & & $3.2 \times 10^{-15}$  & &  \citep{riseberg73:1962,stancu15:045011} \\                
\\
\hline \hline
\end{tabular}
\end{center}
}
\end{table*}

\section{Obelix}
\label{obelixsection}

\subsection{General description}
\label{obelixdescri}

OBELIX 
is a collisional-radiative model with an explicit treatment of the electron kinetics using the Boltzmann equation.
  It was originally developed to model an argon glow discharge \citep{bretagne81:1225,bretagne82:2205,bretagne86:761,bretagne86:779} and then modified to model the ionization region of a dcMS discharge  \citep{guimaraes91:133,guimaraes93:127}. The model is based on solving the Boltzmann equation for the EEDF within the IR.
Recently, OBELIX has been adapted to model a HiPIMS  discharge with argon as the working gas and a copper cathode target \citep{bretagne15:P1.17}. This model was earlier used to study the dependence of the
EEDF on the discharge parameters (discharge current, cathode voltage, working gas pressure) in a dcMS discharge with molybdenum cathode target \citep{guimaraes93:127}. 
The  model has been thoroughly revised for this current study.   This includes adding titanium to the reaction set and revising the argon reactions and cross sections.   In addition, a detailed description of the argon energy levels is added \citep{vlcek89:623}.  The model is  a volume-averaged glow discharge model that takes into account all relevant particle interactions including elastic, inelastic, super-elastic, ionizing and Penning collisions as well as radiative transitions.    


In OBELIX, the electron energy distribution function (EEDF) is calculated self-consistently using the isotropic Boltzmann equation \citep{rockwood73:2348,guimaraes93:127} to which the corresponding collisional terms  are added.  The EEDF is therefore determined by solving the   time-dependent  Boltzmann equation
\begin{equation}
\frac{\partial g_{\rm e}({\cal E}_{\rm e},t)}{\partial t} = - 
 \frac{\partial \Gamma_{\rm e,en}}{\partial {\cal E}_{\rm e}}    - 
 \frac{\partial \Gamma_{\rm e,ee}}{\partial {\cal E}_{\rm e}}   -
 \frac{\partial \Gamma_{\rm e,heat}}{\partial {\cal E}_{\rm e}}   + 
{\cal R}_{\rm e,exc} +  {\cal R}_{\rm e,iz}
 + S - L
\label{tdBol}
\end{equation}
where  $g_{\rm e}({\cal E}_{\rm e},t)  {\rm d} {\cal E}_{\rm e}$ is the number density of electrons having energy in the range $[{\cal E}_{\rm e}, {\cal E}_{\rm e} + {\rm d}{\cal E}_{\rm e}]$ and  $g_{\rm e}({\cal E}_{\rm e},t)$  is  the EEDF  expressed in m$^{-3}$eV$^{-1}$  and $t$ is time. 
The right hand side of Eq.~(\ref{tdBol}) includes the partial derivatives of the electron flux along the energy axis.  The electron flux is divided into separate terms according to the contributing mechanisms. The flux $\Gamma_{\rm e,en}$ is due to elastic electron-atom (electron-molecule) collisions, and $\Gamma_{\rm e,ee}$ is due to electron-electron Coulomb collisions.
Here, an  electron flux term $\Gamma_{\rm e,heat}$ due to Ohmic heating of the electrons is introduced. 
The heating rate 
is imported from the IRM as described below (Section \ref{mergingirmobelix}). The total electron flux along the energy axis $\Gamma_{\rm e,total} = \Gamma_{\rm e,en} + \Gamma_{\rm e,ee} + \Gamma_{\rm e,heat}$ therefore gives an explicit description for  the Boltzmann equation including the effects of Ohmic heating.  
Contributions from inelastic collisions  
that include excitation ${\cal R}_{\rm e,exc}$ and ionization ${\cal R}_{\rm e,iz}$ collisions, are represented by reaction rates per unit energy.  
More details are given by \citet{rockwood73:2348} and/or \citet{bretagne81:1225,bretagne82:2205}.  
 The time-dependent Boltzmann equation (Eq.~(\ref{tdBol})) is discretized using the Rockwood formalism \citep{rockwood73:2348} modified by  \citet{bretagne82:2205}  to incorporate electron energy intervals $w_i$ that grow in width according to a geometrical series with a growth factor $k$ as the electron energy increases
\begin{equation}
w_{i + 1} = k w_i. 
\label{wkw}
\end{equation}
This has the advantage of keeping the code efficient in particular for modeling discharges that exhibit a wide electron energy range such as the HiPIMS discharge.

OBELIX in its current version is not a stand-alone code because it requires input represented by the terms in Eq.~(\ref{tdBol}).  The total electron flux along the energy axis 
gives the explicit form of the Boltzmann equation subjected to a uniform field.  Here, the IRM is used to 
determine $\Gamma_{\rm e,heat}$ for the OBELIX treatment of Eq.~(\ref{tdBol}).  The Ohmic heating is a new feature in this current version of OBELIX. 
In the discretized form, the minimum energy that electrons can gain is the difference between the energy of the initial interval and the energy of the next higher interval. Moreover, the interval widths grow with increasing energy.  The minimum energy gain thus varies depending on which interval the electron comes from. This implies, that not all electrons can be heated by a tiny amount, but rather a few electrons can be heated by the difference between two energy intervals so that the power that goes into the electrons equals the power to Ohmic heating calculated by the IRM.  For that reason, the total power that goes to Ohmic heating is distributed to an appropriate number of electrons that are moved up one energy interval. This is done so that the power to each electron interval is proportional to the number of electrons in that interval. Only the electrons in the last interval remain unheated.

The source term in Eq.~(\ref{tdBol}) is to inject secondary electrons into the IR.  The secondary  electrons emitted from the cathode are accelerated across  the cathode sheath, which is assumed to be non-collisional, and the energy gain is taken to be  roughly $eV_{\rm SH}$ \citep{guimaraes91:133}.    
The source term $S({\cal E}_{\rm e})$ corresponds to the secondary electron current   $I_{\rm see}$ entering  per unit volume 
and is given as
\begin{equation}
S({\cal E}_{\rm e}) = G({\cal E}_{\rm e})I_{\rm see}\frac{1}{e{\cal V}_{\rm IR}}.
\end{equation}
where $G({\cal E}_{\rm e})$ is the energy distribution of the secondary electrons and ${\cal V}_{\rm IR}$ is the volume of the ionization region. 
In this study, the energy distribution $G({\cal E}_{\rm e}) = 1$ at the energy corresponding to $eV_{\rm SH}$, and 0 everywhere else.
The loss term $L$ in Eq.~(\ref{tdBol})
arises from  the application of the quasi-neutrality condition and is therefore dependent upon the ionic diffusion rate (see Section \ref{mergingirmobelix}).


\subsection{The reaction sets for OBELIX}
\label{reactionsetsOBELIX}

The current version of OBELIX uses a simplified argon model described by \citet{vlcek89:632} which has been used by several authors for modeling  argon discharges 
\citep{vlcek89:623,vlcek90:526,bogaerts98:121,bultel02:046406,akatsuka09:043502}. 
This current argon model is based on earlier works by \citet{drawin76rr}, and contains overall 65, individual and effective argon levels, grouped according to their core quantum number $j_{\rm c} = 1/2$ (primed system) and 3/2 (unprimed system). In that model, electron impact excitation cross sections from the ground state and from the excited states are based on semi-empirical cross sections in the form of analytical expressions, given by \citet{drawin67rr} for allowed, parity-forbidden and spin-forbidden transitions.  
The cross sections are calculated using the Born-Bethe formalism \citep{bethe32:293,inokuti71:297},  which is  empirically  modified  at low impacting electron  energy. The electron impact  cross sections among the 65 levels of the argon atom for  energies  ranging  from  thresholds  up  to  the relativistic domain are included in the model.  
  These cross sections are determined from fitting parameters $\alpha_{ij}^{\rm A}$ and $\beta_{ij}$, for transition between levels $i$ and $j$, for optically allowed transitions ($\Delta l = \pm 1, \Delta J = 0, \pm 1$  except $J = 0 \longrightarrow  J = 0$),
\begin{equation}
\sigma_{ij}^{\rm e}(U_{ji}) = 4 \pi a_0^2 \left( \frac{{\cal E}^{\rm H}_{\rm iz}}{{\cal E}_{ji}} \right)^2 
\alpha_{ij}^{\rm A} f_{ij} \frac{U_{ji} - 1}{U_{ji}^2} 
\ln \left( \frac{5}{4} \beta_{ij} U_{ji} \right),
\end{equation}
from fitting parameters $\alpha_{ij}^{\rm P}$ for parity-forbidden transitions ($\Delta l \neq \pm 1$),
\begin{equation}
\sigma_{ij}^{\rm e}(U_{ji}) = 4 \pi a_0^2  \alpha_{ij}^{\rm P}  \frac{U_{ji} - 1}{U_{ji}^2},
\end{equation}
and from fitting parameters $\alpha_{ij}^{\rm S}$  for spin-forbidden transitions  ($\Delta J \neq  0, \pm 1$ including $J = 0 \longrightarrow J = 0$),
\begin{equation}
\sigma_{ij}^{\rm e}(U_{ji}) = 4 \pi a_0^2  \alpha_{ij}^{\rm S}  \frac{U_{ji} - 1}{U_{ji}^2},
\end{equation}
where $U_{ji} = {\cal E}_{\rm e}/{\cal E}_{ji}$ is the reduced kinetic energy of an electron, ${\cal E}_{\rm e}$ is the electron kinetic energy, ${\cal E}_{ji} = {\cal E}_j - {\cal E}_i$, $a_0$ is the first Bohr radius, ${\cal E}^{\rm H}_{\rm iz}$ is the ionization energy of the ground state hydrogen atom, 
and $f_{ij}$ is the absorption oscillator strength \citep{drawin67rr,drawin69:483,bultel02:046406}.  The fitting parameters for the electron impact excitation cross sections are taken from Vl\v{c}ek’s original work \citep{vlcek89:632} with the following exception: For electron impact transitions from ground state to excited states up to the effective level (6d, 8s) (energy level ${\cal E}_i  =$ 15.347 eV) and from the four 4s levels to up to the effective level 5d',7s' (energy level ${\cal E}_i  =$ 15.324 eV), fitting parameters from a recent re-evaluation of the Drawin cross sections are used \citep{bultel02:046406}.   A comparison of electron impact  excitation cross sections to the metastable states from \citet{vlcek89:623} and \citet{bultel02:046406} and the cross section used in the IRM   \citep{yanguasgil05:1588} is given in Appendix \ref{crosssectionappendix}.  Electron impact collisional de-excitation for all forward reactions is considered by applying the principle of detailed balancing \citep[Section 8.5]{lieberman05}.  
Optical de-excitation is considered based on tabulated Einstein coefficients by \citet{drawin76rr}  and we use the adopted tabulated data that was used in Vl\v{c}ek’s original model \citep{vlcek89:623}.     Excitation of Ti to the first 9 excited states are included based on estimated cross sections  as discussed elsewhere \citep{gudmundsson16:065004}, de-excitation cross sections are calculated based on the principle of detailed balancing.

Similarly, electron impact ionization of argon is taken into account from the ground state and the remaining 64 excited states. Following Vl\v{c}ek’s approach \citep{vlcek89:623}, we use the semi-empirical cross sections for argon ionization from level $i$  developed by Drawin \citep{drawin61:513,drawin67rr,drawin69:483}
\begin{equation}
\sigma_{{\rm iz},i}^{\rm e}(U_{{\rm iz},i}) = 4 \pi a_0^2 \left( \frac{{\cal E}^{\rm H}_{\rm iz}}{{\cal E}_{{\rm iz},i}^{\rm Ar}} \right)^2 \xi_i 
\alpha_{i}  \frac{U_{{\rm iz},i} - 1}{U_{{\rm iz},i}^2} 
\ln \left( \frac{5}{4} \beta_{i} U_{{\rm iz},i} \right)
\end{equation}
where  ${\cal E}_{{\rm iz},i} = {\cal E}_{\rm iz} - {\cal E}_i$ and $U_{{\rm iz},i} = {\cal E}_{\rm e}/{\cal E}_{{\rm iz},i}$,  
using the fitting parameters $\alpha_i$ and $\beta_i$ from  \citet{vlcek89:623} for $1 \leq i \leq 11$, where $i = 1$ is the argon ground state, and $\xi_i$ is the number of energetically
equivalent electrons in shell $i$  and $\xi_i = 6$ for $i=1$ and $\xi_i =1$ for $i>1$.  For $i \geq 12$ the values $\alpha_i = 0.67$ and $\beta_i = 1$ are used.   
   A comparison of electron impact ionization cross sections used in OBELIX from \citet{vlcek89:623} and the electron impact ionization from the ground state argon atom experimentally determined by   \citet{straub95:1115} and from the metastable argon states experimentally determined by \citet{dixon73:405} are shown in Appendix \ref{crosssectionappendix}.  
Note that since $l = 1$ and $s = 1/2$ for the last  electron, the core angular momentum
may have two possible values: $j_{\rm c} =  l + s = 3/2$ and $j_{\rm c} = l - s = 1/2$.  Therefore, the two ionization limits have slightly different energies. 
The first core configuration ($j_{\rm c} = 3/2$), also referred to as  'nonprimed' subsystem, has an ionization limit ${\cal E}_{\rm iz} = 15.760$ eV and the second one ($j_{\rm c} = 1/2$), referred to as the 'primed' subsystem has an ionization limit ${\cal E}'_{\rm iz} = 15.937$ eV \citep{drawin76rr}. 
Consequently, the ionization energy threshold from both the metastable levels is 4.21 eV.

Elastic electron-neutral collisions are taken into account using the formalism developed for non-equal electron energy intervals by \citet{bretagne82:2205} which is based on the work of \citet{rockwood73:2348}.  The elastic electron-Ar cross section used here is a fit given by \citet{bretagne82:2205} to experimental data from  \citet{frost64:A1538}.   

The cross section for electron impact ionization of the titanium atom from the ground state is based on the cross sections from  \citet{bartlett04:235}, and for the electron impact ionization reaction  e + Ti$^+$  ~$\rightarrow$~ Ti$^{2+}$ + 2e, the cross section is taken from   \citet{diserens88:2129}. 
The cross sections for Penning ionization  Ar(4s'[1/2]$_0$)  
+ Ti  ~$\rightarrow$~  Ar + Ti$^+$ + e  
and  Ar(4s[3/2]$_2$)  
+ Ti  ~$\rightarrow$~  Ar + Ti$^+$ + e  are taken into account using a constant rate coefficient of $3.2 \times 10^{-15}$  m$^3$s$^{-1}$ as discussed by \citet{stancu15:045011} and based on \citet{riseberg73:1962}.                 
Incorporation of electron impact excitation of Ti is included based on 
earlier work \citep{gudmundsson16:065004}. 
Electron impact excitation of Ar$^+$ is neglected as the densities  remain more than one order of magnitude below the Ar ground state density as discussed in Section  \ref{comparison}. 
Nonresonant charge transfer  Ar$^{\rm +}$ + Ti   ~$\rightarrow$~  Ar + Ti$^+$  is included using a rate coefficient $1 \times 10^{-15}$ m$^3$s$^{-1}$  \citep{gudmundsson16:065004}.

\subsection{Merging of IRM and OBELIX}
\label{mergingirmobelix}

\begin{figure}
\begin{center}
\includegraphics[scale=0.07]{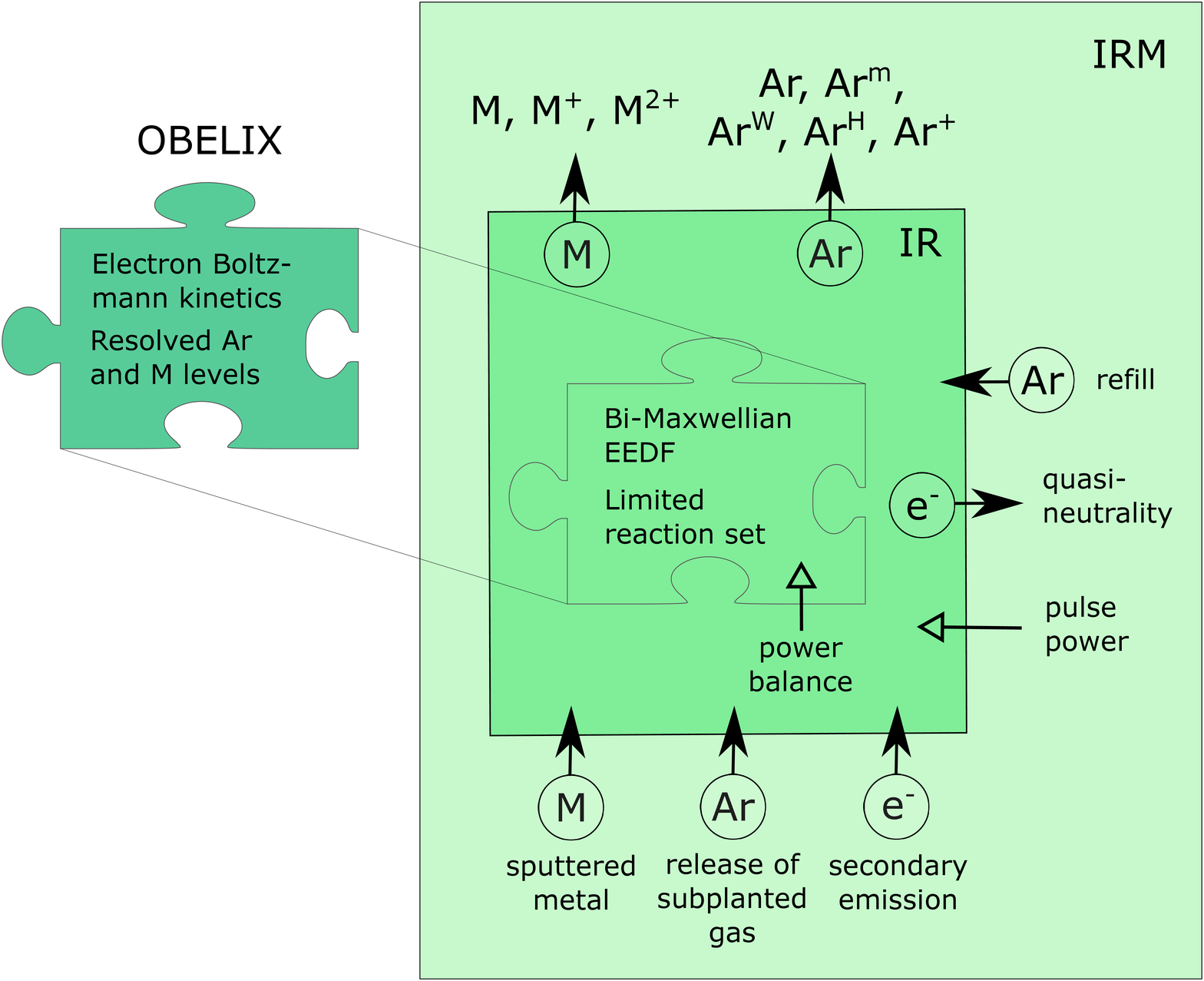}
\end{center}
 \caption{Graphical representation of the IR modeled by the merged IRM and OBELIX model. When using OBELIX, the bi-Maxwellian electron energy distribution in IRM is replaced by an explicit treatment of the electron kinetics. In addition, the energy cost of ionization-concept of the IRM model is replaced by representing the neutral Ar atom with 65 individual and effective levels and the neutral Ti level by one ground state and 9 excited levels. Arrows indicate sputter, diffusion and kick-out processes in and out of the IR. 
Note that the combined metastable levels
(Ar(4s'[1/2]$_0$) 
+ Ar(4s[3/2]$_2$)) 
are denoted by Ar$^{\rm m}$.
 \label{merging}}
\end{figure}

In the current version, the OBELIX model is merged with
the IRM 
(Section \ref{irmsection}).  
In that way, the strengths of each model are combined: OBELIX provides an exact treatment of the electron kinetics, while the IRM provides the magnetron sputtering discharge-specific effects and mechanisms, including plasma surface interactions and electron power absorption mechanisms.  The merging of the IRM and OBELIX is shown schematically in Figure \ref{merging}.  The main change by incorporating OBELIX is that the rate coefficients used in IRM are replaced by rate coefficients that are calculated using the EEDF that is obtained by solving the Boltzmann equation.  At the same time a cross-calibration of the models is possible.  This includes exploring the  assumptions and approximations, such as, but not limited to, the approximation of the electron energy distribution function by two Maxwellian distributions applied in the IRM \citep{huo17:354003} and its influence on the temporal development of the particle densities. 

For the merged model, we take the particle fluxes in and out of the IR from the IRM and use these as input to OBELIX.  This includes the Ar$^+$, Ti$^+$ and Ti$^{2+}$ ion flux out of the IR towards the cathode and the flux of neutral atoms, Ti, 
Ar$^{\rm W}$ and Ar$^{\rm H}$ as well as secondary electrons into the IR from the cathode. Here, Ar$^{\rm H}$ denotes  hot  argon  atoms 
in the ground state, which return from the target immediately after the argon ion impact event, with a typical sputter energy of a few eV,
and Ar$^{\rm W}$ denotes  warm argon atoms in the ground state that are assumed to be embedded in the target at the location of ion impact, and then return to the surface and leave with the target temperature, at most  0.1 eV \citep{huo14:025017,gudmundsson16:065004}. 
Out-diffusion of film-forming species (Ti, Ti$^+$ and Ti$^{2+}$) and Ar working gas species (Ar, Ar$^+$, Ar$^{\rm H}$, Ar$^{\rm W}$, Ar(4s'[1/2]$_0$), Ar(4s[3/2]$_2$)) are similarly taken from the IRM.  
The combined metastable levels (Ar(4s'[1/2]$_0$) + Ar(4s[3/2]$_2$)) are denoted by Ar$^{\rm m}$. Finally, Ar diffusion into the IR to refill the volume is also taken from the IRM.  The electron diffusion rate out of the IR is calculated within OBELIX to obey charge neutrality.
Cross sections for volume reactions are then used to calculate the electron impact rate coefficients and the  reaction rates as the electron energy distribution function is known. 
The volume reaction rates, i.e.~excitation and ionization of Ar and Ti are calculated by OBELIX based on the heavy species densities and the EEDF at that moment within the pulse. 
 These reaction rates therefore determine the temporal variations of the various species.
Penning collisions and charge exchange collisions are taken into account based on the heavy species densities in the OBELIX volume.
The volume reaction rates, including the electron-electron interaction, are treated in OBELIX.  Energization of electrons is determined by taking the energy flux of each of the two electron power absorption channels from the IRM. These time-varying fluxes are injected into the OBELIX discharge model  at the correct moment in the pulse. 
The total power that goes to Ohmic heating is distributed over the energy intervals proportional to the electron density in each interval. The electron flux $\Gamma_{\rm e,heat}$ is then determined by the number of electrons that are moved up one energy interval using that part of the distributed total power of Ohmic heating that is deposited in the energy interval. 
The numerical procedure implements the additional term  $\partial \Gamma_{\rm e,heat}/\partial {\cal E}_{\rm e}$ in Eq.~(\ref{tdBol}) for the electron flux in energy due to Ohmic heating and is described in Section \ref{obelixdescri}. 
A comparison is made between the results from the IRM and the results from the merged IRM and OBELIX.  The latter we refer to  simply as results from OBELIX.



\section{Results and discussion}
\label{resultsanddiscussion}

\subsection{Discharges and model inputs}

\begin{figure}
\begin{center}
\includegraphics[scale=0.34]{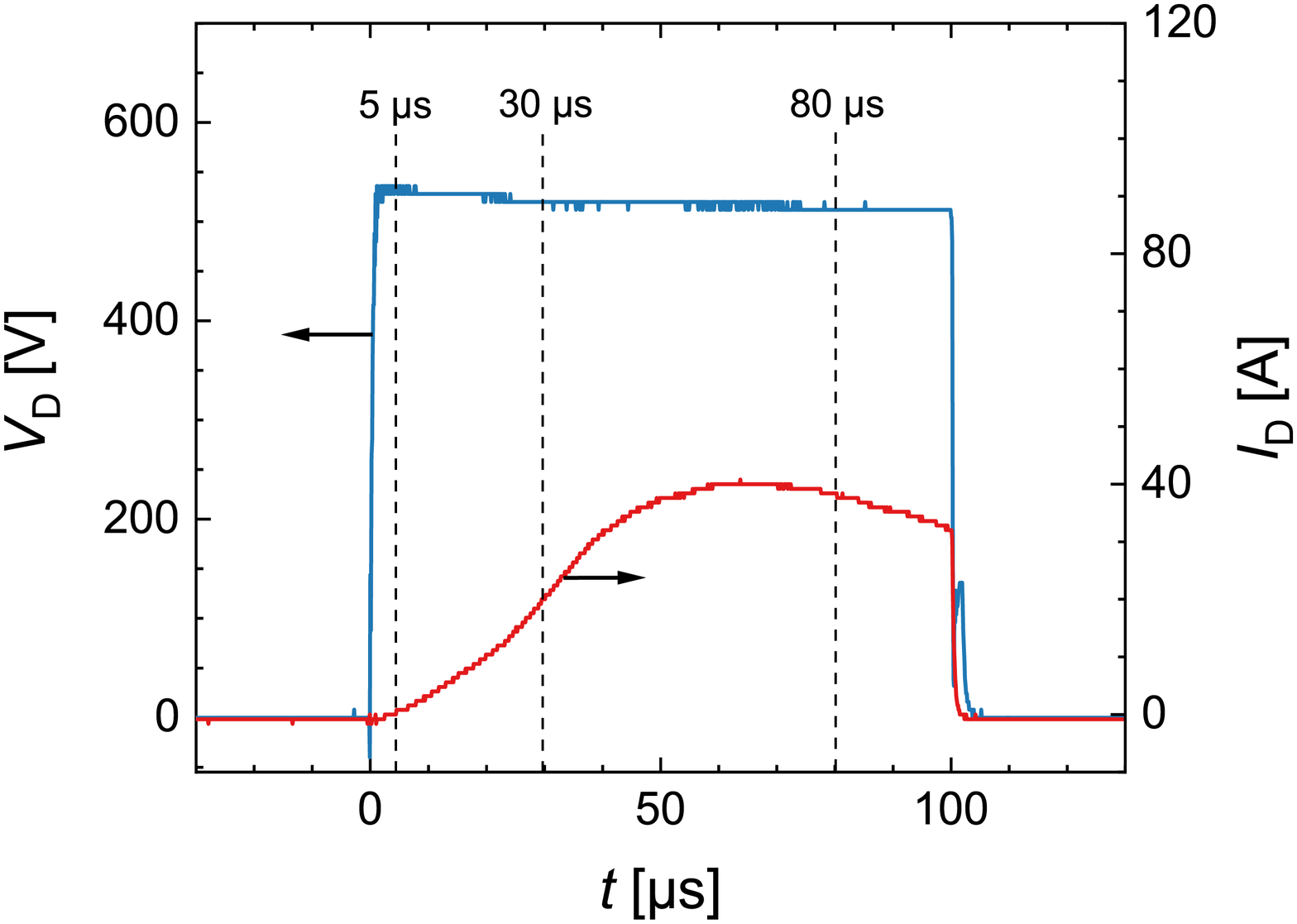}
\end{center}
 \caption{The measured discharge current and voltage waveforms recorded for the magnet configuration C0E0
by \citet{hajihoseini19:201}.  The argon working gas pressure was set to 1 Pa and the cathode target was made of titanium and 4 inch in diameter.
The pulse width was 100 $\mu$s and the average power 300 W.  The dashed vertical lines indicate the moments at which the EEPFs in Figure \ref{eedffig} are recorded.
\label{idvd}}
\end{figure}

To investigate the EEDF in a HiPIMS discharge 
 and compare the assumed EEDF used in the IRM to the EEDF calculated self-consistently in the OBELIX model, we select one of the 
discharges that was analyzed experimentally by \citet{hajihoseini19:201,hajihoseini20:033009} and was already modeled using the IRM by \citet{rudolph20:05LT01,rudolph21:033303}.  This  discharge was operated with argon  working gas at 1 Pa using a 4 inch (102 mm diameter)  Ti cathode target.
The ionized flux fraction $F_{\rm flux}$ and the deposition rate were measured at 30 mm above the race track using a grid-less ion meter \citep{kubart14:152}.  For details on the experiments and the magnetic field topology, see \citet{hajihoseini19:201}.
The discharge  was generated using a magnetic field configuration denoted C0E0 \citep{hajihoseini19:201}, which indicated that both the center and edge magnets sit next to the back of the cathode target and give the highest magnetic field strength (parallel to the target surface) in the cathode target vicinity just above the target race track.   
The time-averaged power to the discharge was maintained at 300 W, the pulse length 100 $\mu$s, and the pulse repetition frequency was $f = 54$ Hz.  The measured discharge current and voltage waveforms recorded for this magnet configuration are shown in Figure \ref{idvd}.

For the present work, the IRM is updated after a thorough re-evaluation of cross sections, which leads to the substitution of cross sections involving the  two metastable levels, reactions (\ref{Ra:2}) -- (\ref{Ra:12}) in Table \ref{ratecoeff} (see Section  \ref{reactionsetsIRM}). The revised model is  run using a well-established fitting procedure 
\citep{huo17:354003,butler18:105005}.  This gives  slightly different adjustable parameters for  the  potential   $V_{\rm IR}$ that  drops  over the IR and the target ion back-attraction factor $\beta_{\rm t,pulse}$, compared to the values published earlier by \citet{rudolph20:05LT01}.    Table \ref{parametertable}  summarizes the measured values and the IRM fitting parameters relevant for this current study. Similar to the earlier studies by \citet{rudolph20:05LT01,rudolph21:033303}, the electron recapture probability $r$ is set to  0.7. 

The particle fluxes in and out of the IR from the constrained IRM are  then used as an input to OBELIX, where the volume reaction rates are  recalculated using the explicit treatment of the electron kinetics, as well as the temporal evolution of the particle densities. The EEDF is discretized using non-equal energy intervals \citep{bretagne82:2205}  between 0 eV and a value above the energy that corresponds to the sheath voltage $eV_{\rm SH}$.  The secondary electrons are injected   into the discharge volume at an energy close to the energy corresponding to the sheath voltage $V_{\rm SH} = V_{\rm D} \times (1-V_{\rm IR}/V_{\rm D})$. 
The cathode potential is 510 V  for the peak discharge current of 41 A  \citep{hajihoseini19:201}.
 The sheath voltage is therefore $V_{\rm SH}  = 510$ V $\times \ (1 - 0.099) = 460$ V. 
The energy interval width is 1 eV for the first electron energy interval (0 to 1 eV) and grows larger with increasing energy according to Eq.~(\ref{wkw}), until it reaches a width of close to 2 eV for the last energy interval at the highest energy considered.

To initiate the discharge (at $t = 0$ s), the IRM requires the presence of some seed charge carriers. 
The initial conditions for the cold electron temperature and density are chosen $T_{\rm ec,0}$ = 0.5 eV and $n_{\rm ec,0}$ = 10$^{16}$ m$^{-3}$, respectively. The initial hot electron density is chosen to be $n_{\rm eh,0}$ = 10$^{3}$ m$^{-3}$. 
For OBELIX a similar seed charge density is chosen.  The initial cold electron population is a Maxwellian distribution with a density of 10$^{16}$ m$^{-3}$ and electron temperature of 0.5 eV, while no hot component is included initially.   Note that the shape of the initial EEDF has no influence on the shape of the EEDF at a later stage.  However, it has an influence on the computational time, which is why we choose a Maxwellian distribution as an initial EEDF.
\begin{table}
\caption{The measured discharge parameters from \citet{hajihoseini19:201}, the IRM fitting parameters \citep{rudolph21:033303}, and the energy discretization parameters for OBELIX.  \label{parametertable}}
\begin{center}  
\begin{tabular}{ll}
\hline \hline
&   \\ 
IRM &  \\   
\hline
$I_{\rm D, peak}$ [A] &  41   \\  
$V_{\rm D}$ [V] &  510  \\
$V_{\rm IR}/V_{\rm D}$ & 0.099 \\
$\beta_{\rm t,pulse}$ &  0.89 \\ 
$F_{\rm flux}$  & 0.17   \\
  & \\
OBELIX & \\
\hline
${\cal E}_{\rm e}$ [eV] & 0 -- 481   \\
number of intervals & 340 \\
smallest interval energy  [eV] & 1   \\
initial $n_{\rm e0}$ [m$^{-3}$] & $10^{16}$ \\
initial $T_{\rm e0}$ [eV] & 0.5  \\
&   \\
\hline \hline
\end{tabular}
\end{center}

\end{table}

In the merged model, OBELIX calculates the volume reaction rates while the fluxes in and out of  the IR are taken from the IRM. Flux here refers  to diffusion, kick-out, and sputtering involving the species Ti, Ar$^{\rm H}$, Ar$^{\rm W}$, Ar$^{\rm m}$, Ar$^+$, Ti$^+$ and Ti$^{2+}$.   This could lead to an accumulation or depletion of certain species from slightly different volume production rates in OBELIX and the IRM, that could, over time, lead to high differences in density. In order to avoid this, for the merged model, we adopt the flux from the IRM according to 
\begin{equation}
R_{k,{\rm OBELIX}}(t) = \frac{n_{k,{\rm OBELIX}}(t)}{n_{k,{\rm IRM}}(t)} R_{k,{\rm IRM}}(t)
\label{rktleid}
\end{equation}   
where $R_k(t)$ is the flux of species $k$ at time $t$ out of or into the IR
\citep{huo17:354003} and $n_k(t)$ is the volume-averaged density of species $k$ at time $t$. 
The adaption of the diffusion and kick-out rates out of the IR is only a small adoption. This can be seen from Figure \ref{denscompar} which shows that  the OBELIX densities remain within an interval of $\pm 40$ \% of the IRM densities at the end of the pulse.

Excited Ar and excited Ti species are not explicitly modeled in the IRM. Their diffusion rates and kick-out rates are calculated according to
\begin{equation}
R_{k,{\rm OBELIX}} = \frac{n_{k,{\rm OBELIX}}(t)}{n_{l,{\rm IRM}}(t)}  R_{l,{\rm IRM}} 
\end{equation}  
where $n_{l,{\rm IRM}}(t)$ and $R_{l,{\rm IRM}}$ are the density and diffusion rate of an excited species from the IRM (e.g.~Ar(4s'[1/2]$_0$) etc.), respectively, and $k$ stands for excited species of argon or excited species of titanium from OBELIX.

\subsection{Comparison between the IRM and the OBELIX model}
\label{comparison}

\begin{figure}
\begin{center}
\includegraphics[scale=0.36]{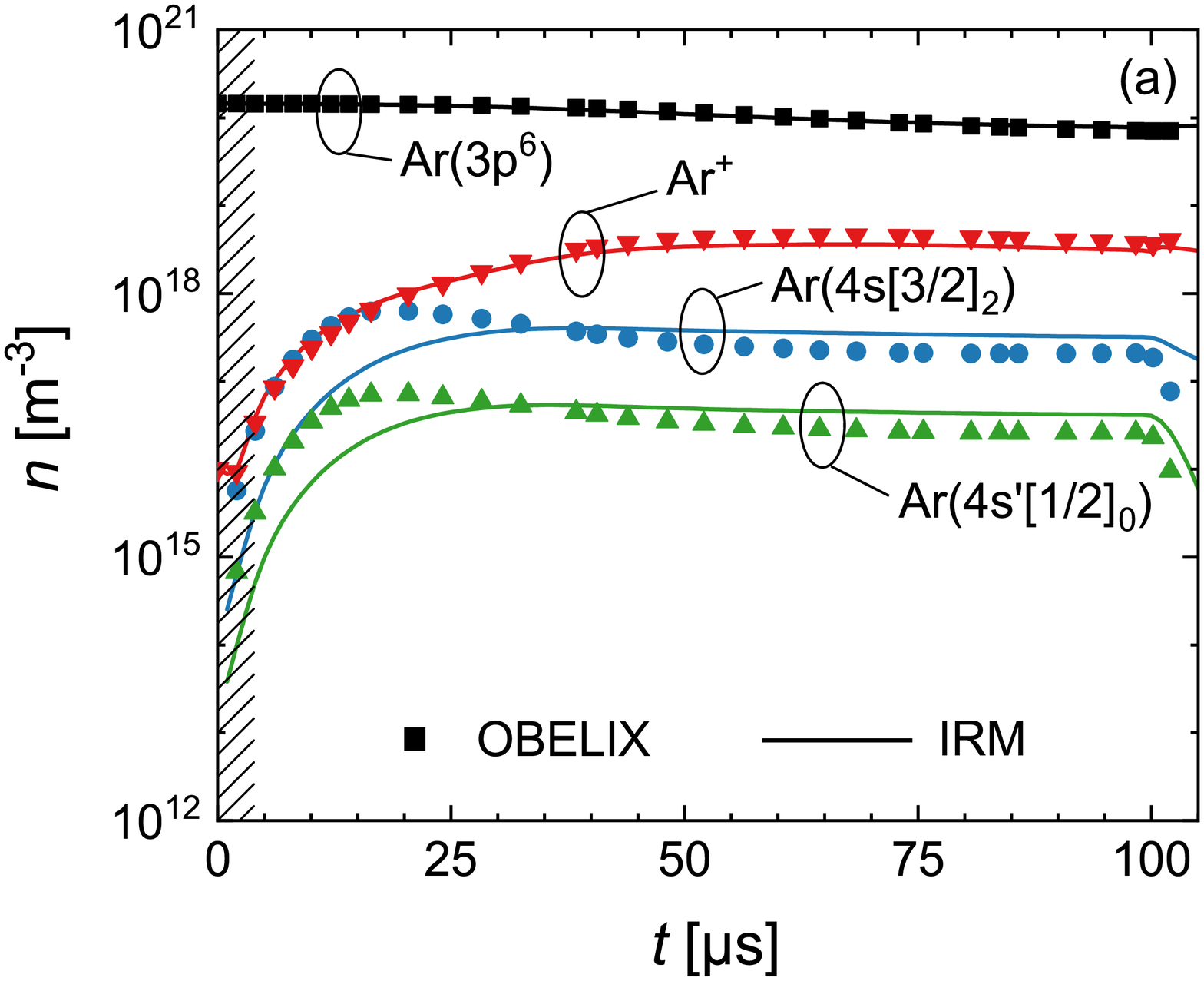}
\includegraphics[scale=0.36]{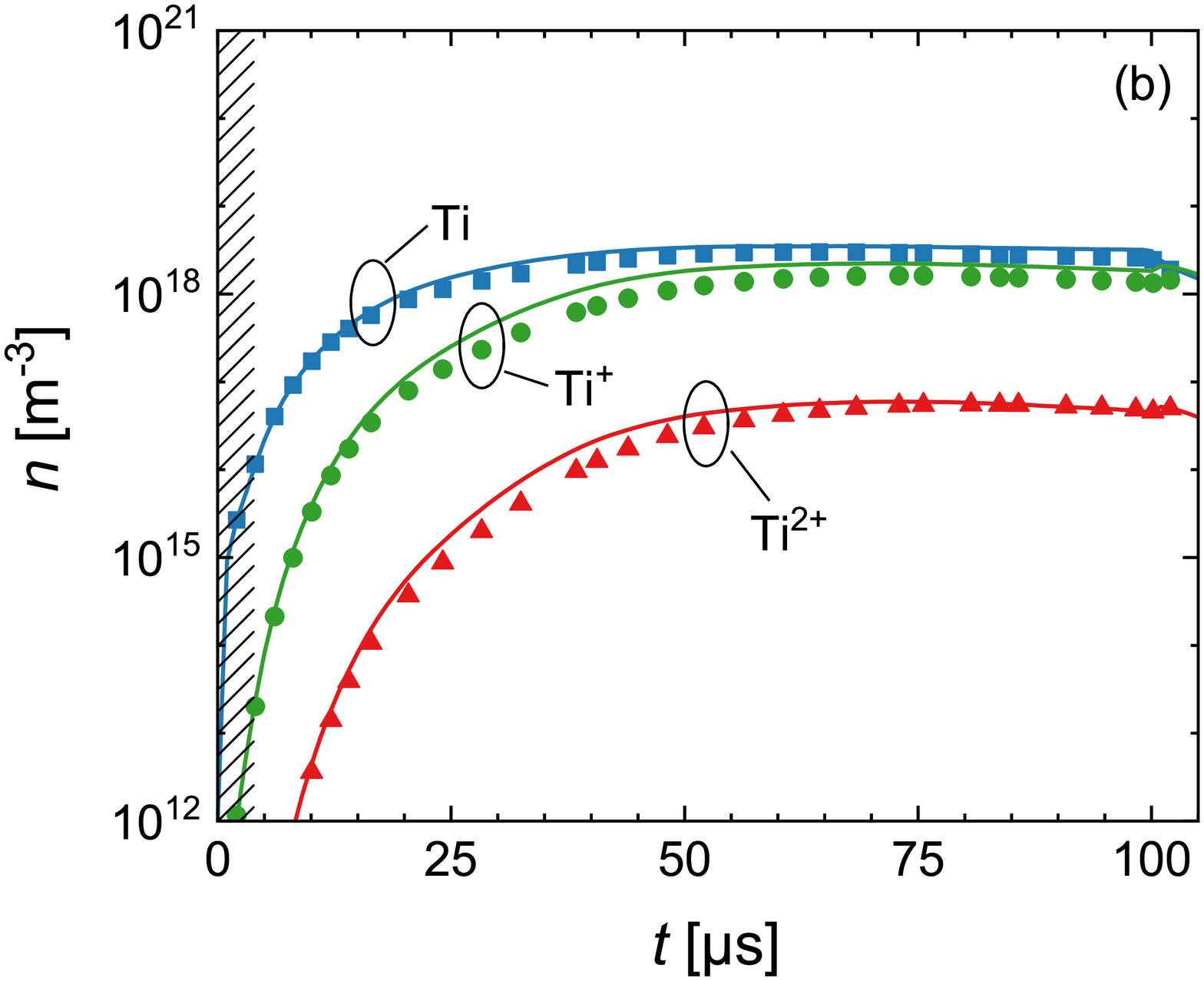}
\end{center}
 \caption{The temporal density evolution of the principal (a) argon working gas species and  (b) sputtered titanium species  during a pulse calculated from the IRM (full lines)  and OBELIX (symbols) for a discharge with 4 inch titanium target operated at 1 Pa with  a peak current of $I_{\rm D,peak}$ = 41 A.  The hatched area at the beginning of the pulse ($t <$ 5 $\mu$s) indicates a lack of precision for both models. 
\label{denscompar}}
\end{figure}

Figure \ref{denscompar} (a) shows the temporal evolution of the density of the argon working gas species Ar, Ar(4s[3/2]$_2$), Ar(4s'[1/2]$_0$), and Ar$^+$,  during the discharge pulse, calculated by the IRM and by the OBELIX model.  The first four microseconds are dashed in the figure to indicate uncertain data. This data is questionable as the OBELIX results depend on the IRM, which is believed to not yield reliable results at an early stage of the pulse. This is due to the 
seed density of electrons at the pulse start which is required for running the IRM. The seed density remains a significant part of
the modeled electron density in the initial stages of the pulse. By varying the seed density  in the range $10^{15} - 10^{16}$ m$^{-3}$ we find that the model has converged at around 4 $\mu$s in the sense that the result is independent of the seed value. Therefore, the first few $\mu$s are uncertain (the hatched area in Figures \ref{denscompar}, \ref{ratefig} and \ref{coifig}) (see also \citet{raadu11:065007}).
In general, there is a very good match for the temporal behavior of all species between the two models.  However,  the argon metastable densities exhibits a much faster rise in the beginning of the pulse as well as  a slightly faster drop towards the end of the pulse  in the OBELIX results compared to  the IRM results.  Keep in mind that the treatment of the excited levels is significantly more detailed in the OBELIX than in the IRM.  
 Figure \ref{denscompar} (b) shows the temporal evolution of the density of the  sputtered  species Ti, Ti$^+$, and Ti$^{2+}$.    The comparison between the results from IRM and OBELIX reveals an excellent match for  most of the species, except that the Ti$^+$ and Ti$^{2+}$ ion densities are slightly lower in the OBELIX  results 20 -- 70 $\mu$s into the pulse.

\begin{figure}
\begin{center}
\includegraphics[scale=0.36]{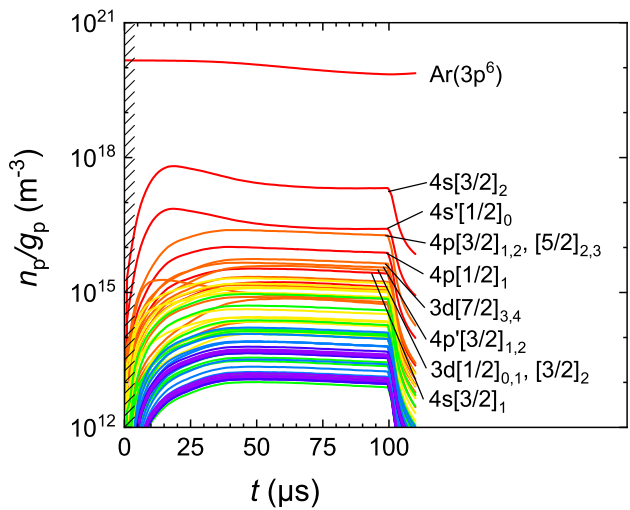}
\end{center}
 \caption{The reduced population density evolution of the Ar excited species from the OBELIX calculations for a discharge with 4 inch titanium target operated at 1 Pa with peak discharge current  of  $I_{\rm D,peak}$ = 41 A. 
\label{arexctime}}
\end{figure}

Figure \ref{arexctime} shows the temporal evolution of the population densities of the excited argon species for the discharge 
calculated by the OBELIX model. The densities of all the levels is normalized by their statistical weight.
 Therefore, this is often referred to as the reduced population density.
The fastest rise in densities is experienced by the two metastable levels Ar(4s[3/2]$_2$) and Ar(4s'[1/2]$_0$) as found earlier experimentally using tunable diode-laser absorption spectroscopy (TD-LAS)  \citep{stancu15:045011}. 
At the beginning of the pulse, the electron density is still low, so both argon metastable levels lack a loss channel at this stage of the pulse. As the electron density rises, electron impact de-excitation (super-elastic collision) reduces the argon metastable density after an initial peak at around 10 $\mu$s. 
The highest population densities  are observed for the  argon metastable levels 4s[3/2]$_2$ and
4s'[1/2]$_0$. They are followed by the (4p[3/2]$_{1,2}$ + 4p[5/2]$_{2,3}$) and the  4p[1/2]$_1$ levels.  This is no surprise as the effective level contains two individual levels that have  quasi-metastable character (mostly transitions to the metastable 4s levels) and the individually modeled level is a quasi-metastable level as well \citep{katsonis11:896836}.

\begin{figure*}
\begin{center}
\includegraphics[scale=0.29]{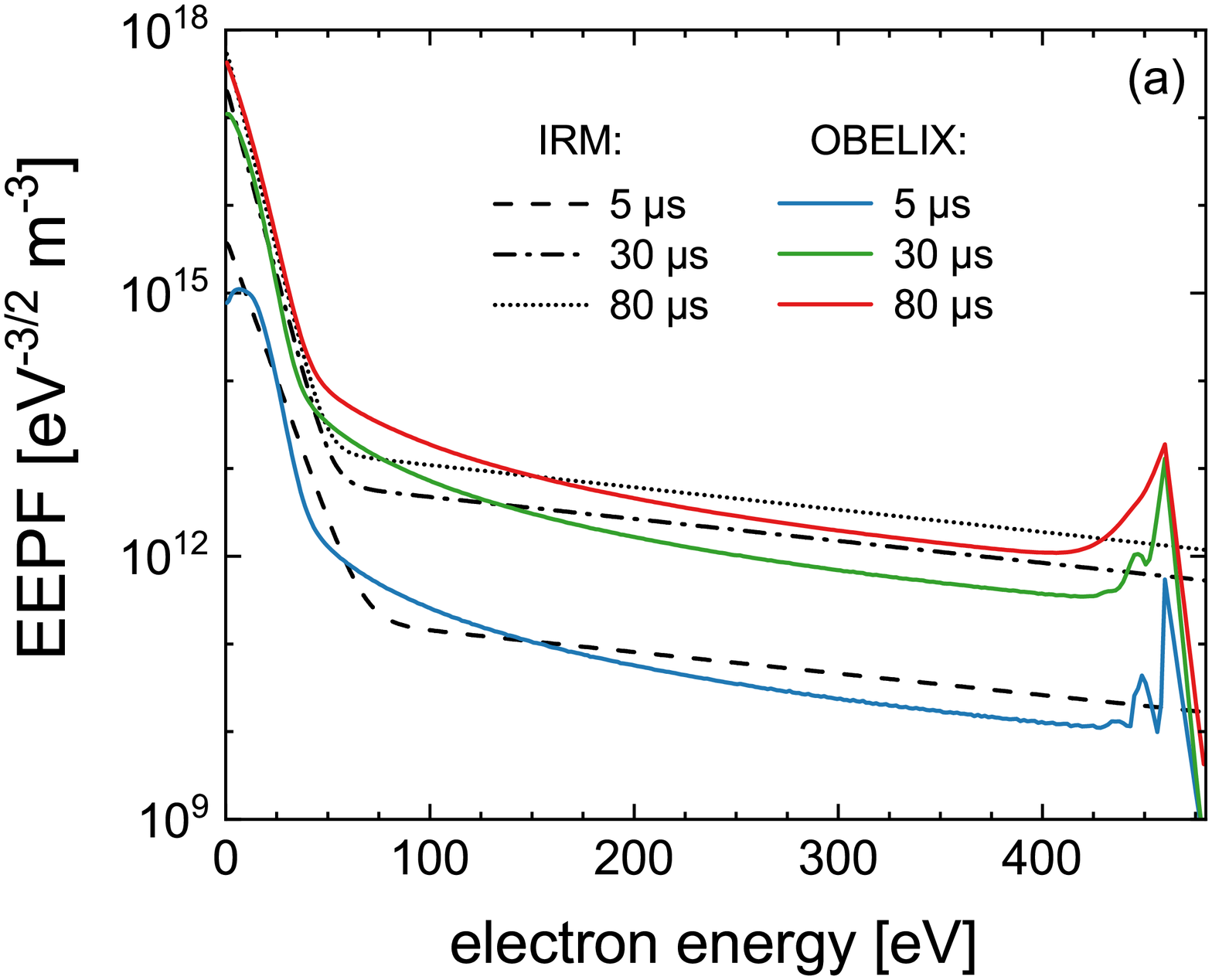}
\includegraphics[scale=0.29]{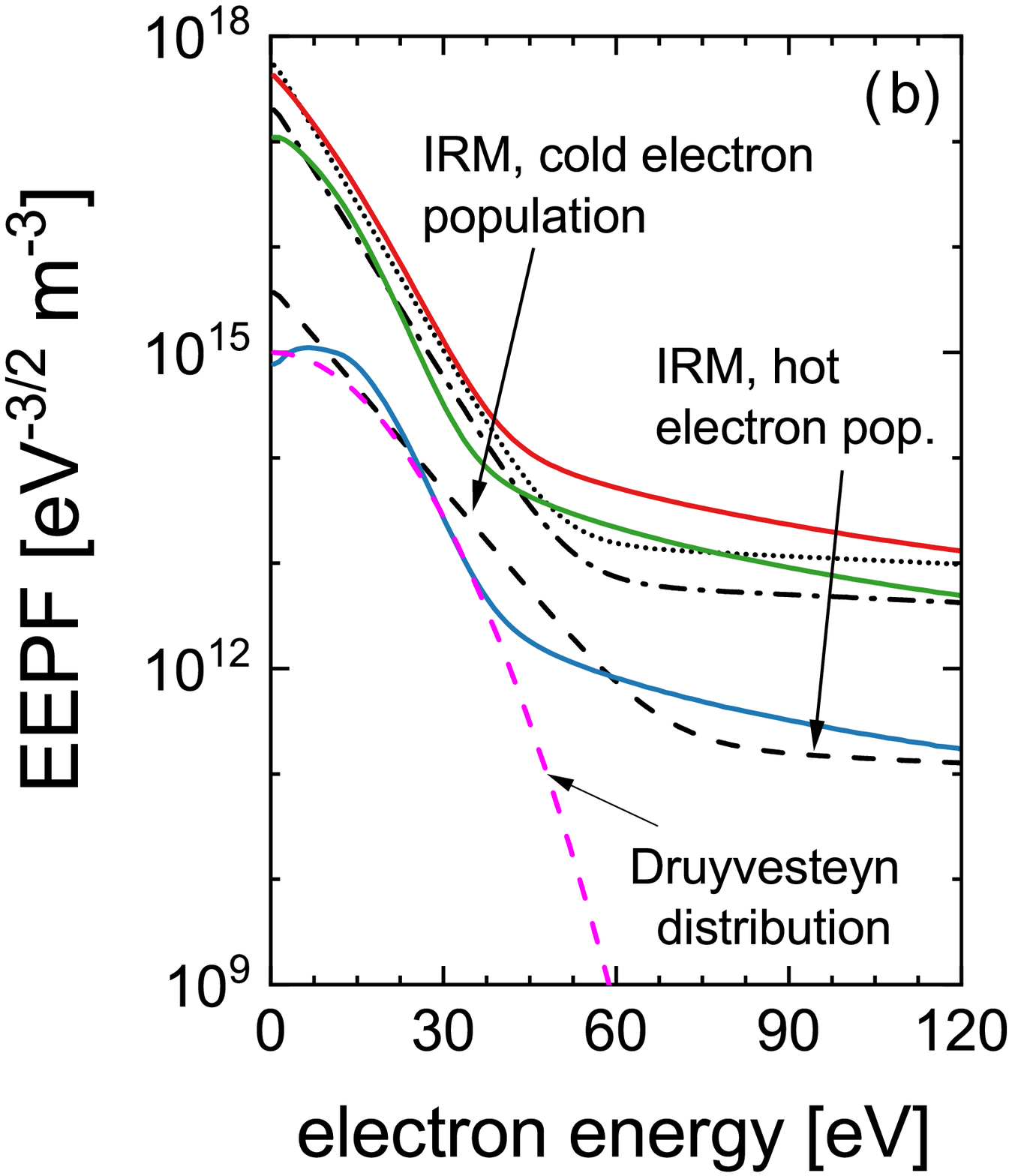}
\includegraphics[scale=0.29]{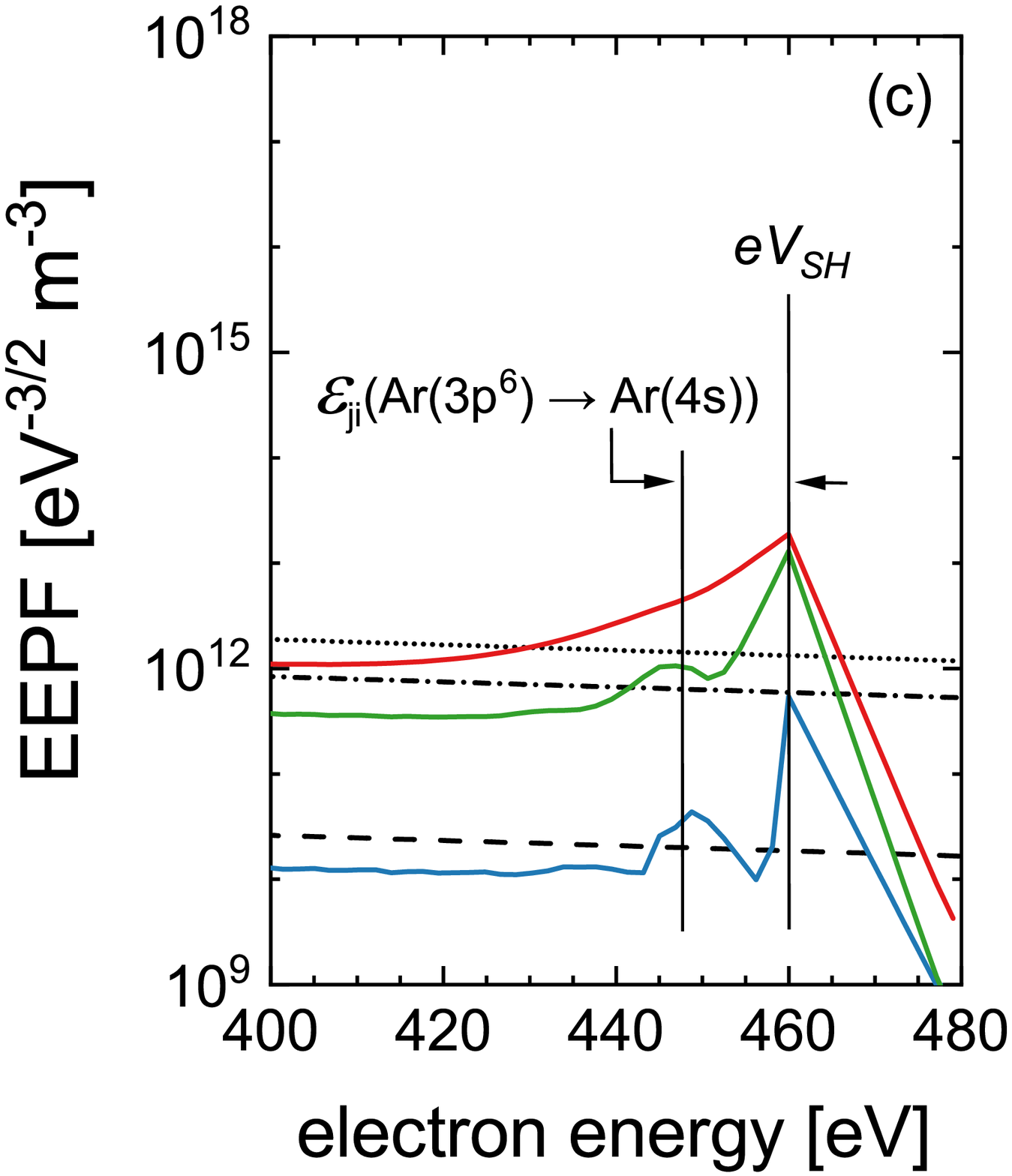}
\end{center}
 \caption{The electron energy probability function (EEPF) at different times in the discharge pulse (pulse initiation (5 $\mu$s), current rise  (30 $\mu$s), and plateau region  (80 $\mu$s))  for a discharge with a 4 inch titanium target operated at 1 Pa with peak discharge current  of  $I_{\rm D,peak}$ = 41 A.  The  (a) full energy range, (b) the low-energy electron range, and (c) the high-energy electron range.   Note that the combined 4s levels (Ar(4s[3/2]$_2$), Ar(4s[3/2]$_1$), Ar(4s'[1/2]$_0$) and Ar(4s'[1/2]$_1$)) are denoted by Ar(4s). 
\label{eedffig}}
\end{figure*}

\begin{figure}
\begin{center}
\includegraphics[scale=0.3]{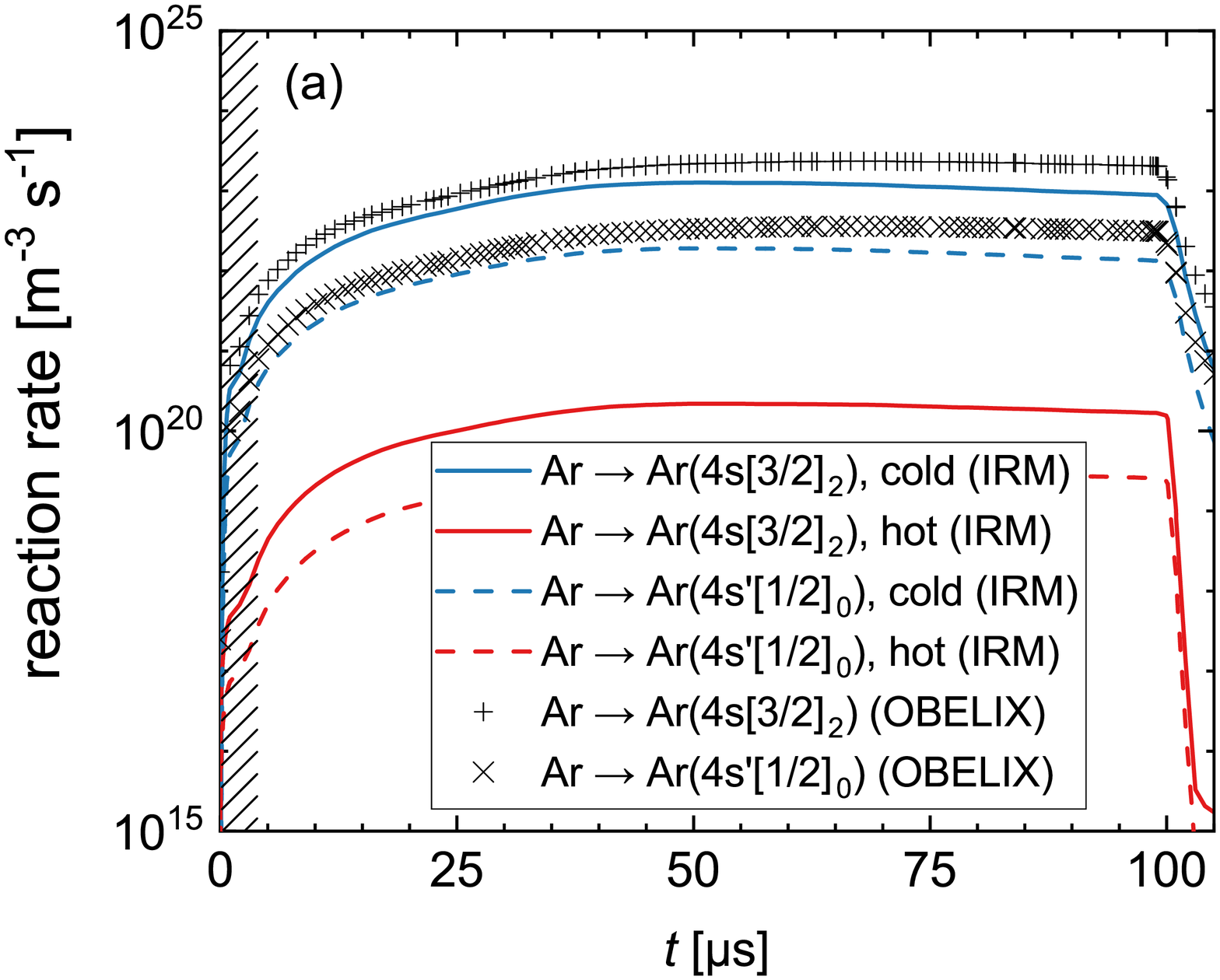}
\includegraphics[scale=0.3]{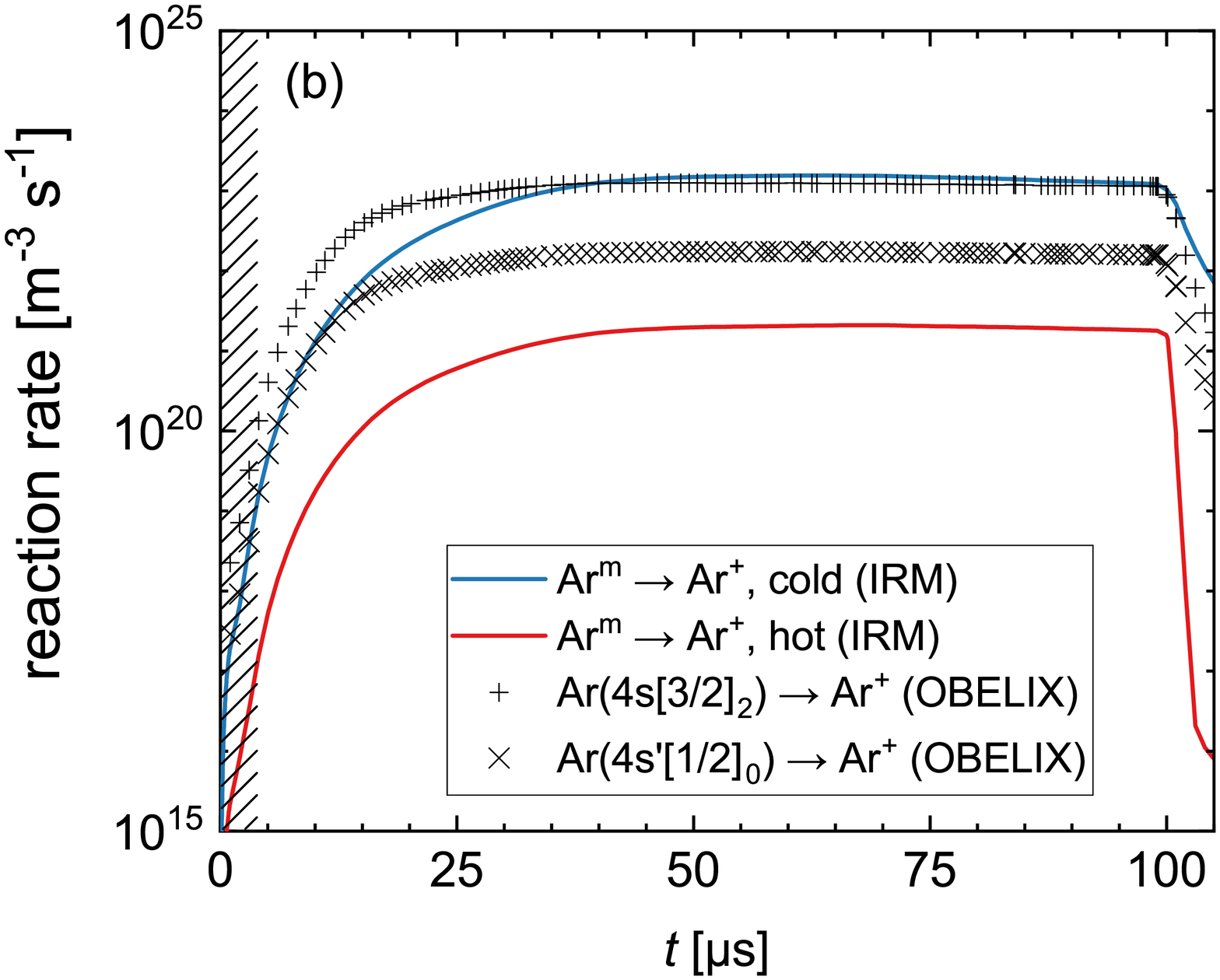}
\includegraphics[scale=0.3]{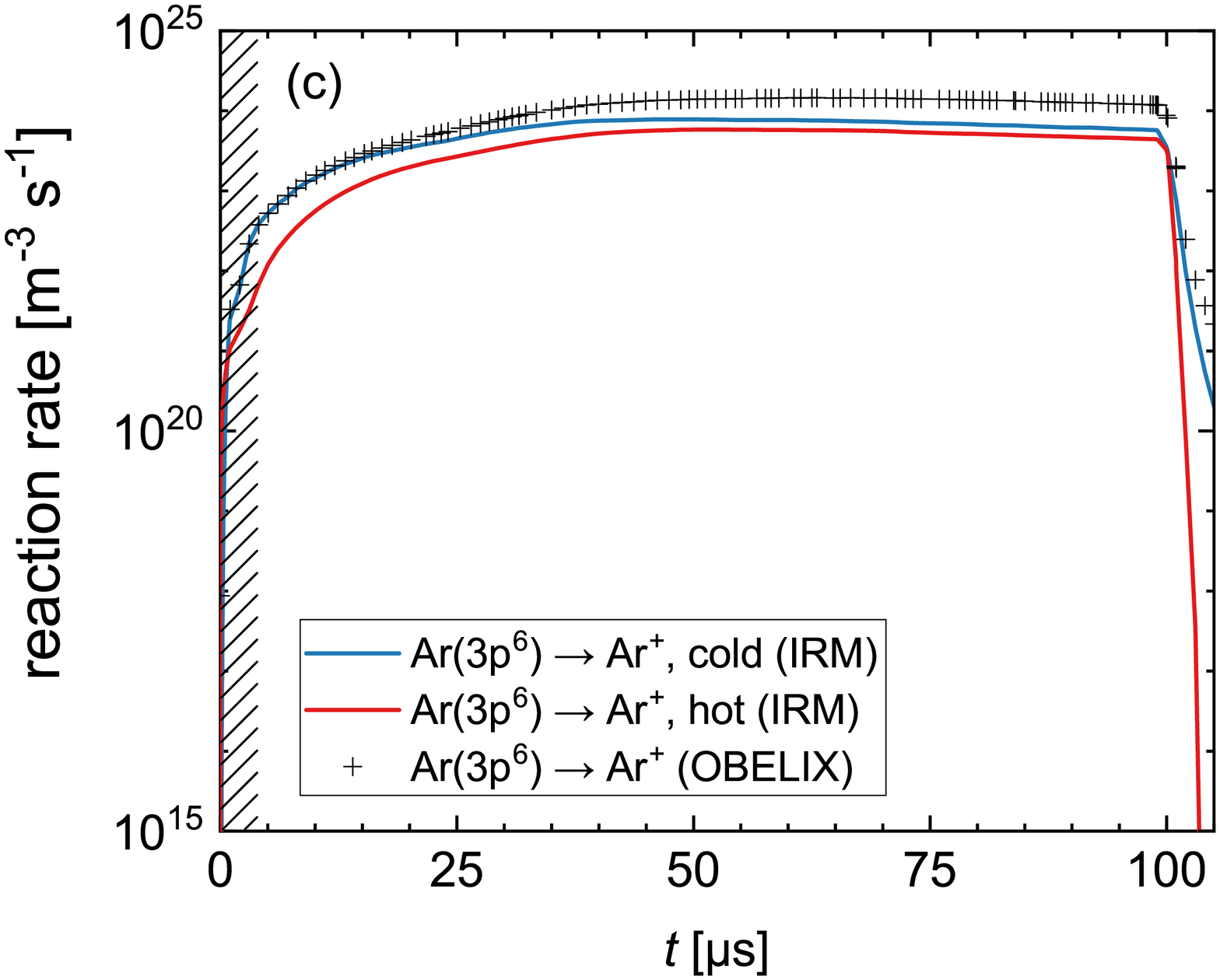}
\includegraphics[scale=0.3]{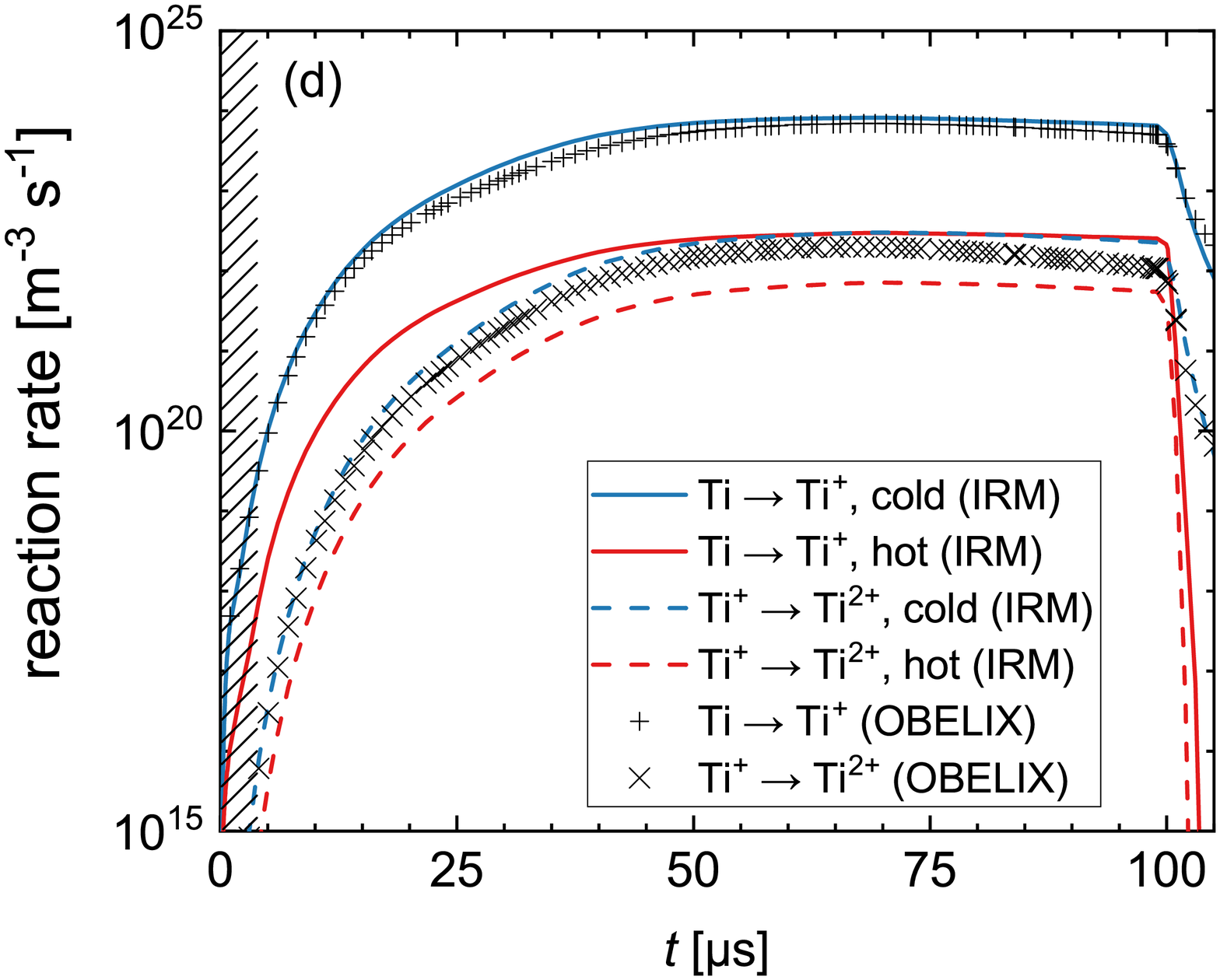}
\end{center}
 \caption{The temporal variation of the reaction rates  (a) for electron impact excitations to the metastable argon level from ground state, (b) for electron impact ionization from the metastable argon levels, (c) for electron impact ionization from the ground state argon atom and (d) for electron impact ionization of the titanium atom and Ti$^+$ to create Ti$^{2+}$. The panels compare the reaction rates used by the IRM (full and dashed lines) for hot and cold electrons and the rate coefficients calculated by OBELIX (crosses). Note
that the combined metastable levels (Ar(4s[3/2]$_2$) and Ar(4s'[1/2]$_0$)) are denoted by Ar$^{\rm m}$.
\label{ratefig}}
\end{figure}

Figure \ref{eedffig}  shows the electron energy probability functions (EEPFs) determined by OBELIX and calculated based on the assumption of a bi-Maxwellian distribution in the IRM.  
The EEPF is defined as $g_{\rm p}({\cal E}_{\rm e}) = {\cal E}_{\rm e}^{-1/2}  g_{\rm e} ({\cal E}_{\rm e})$ where $ g_{\rm e} ({\cal E}_{\rm e})$ is the EEDF.  For a Maxwellian distribution  $\ln(g_{\rm p}({\cal E}_{\rm e}))$ is linear with ${\cal E}_{\rm e}$. 
The EEPF is shown at three different times 
during the pulse: ignition ($t$ = 5 $\mu$s), current rise ($t$ = 30 $\mu$s), and finally close to the end of the pulse ($t$ = 80 $\mu$s), as indicated on the discharge current and voltage waveforms in Figure \ref{idvd}.  
The EEPF absolute value increases, with increased electron density as the pulse evolves. Most of the electrons can be found at low energy,
this is where the hot electrons pile up after they have lost most of their energy in both elastic and inelastic collisions. 
At 5 $\mu$s into the pulse, the low energy part of the EEPF determined by OBELIX exhibits a Druyvesteyn-like distribution, which is an indication of elastic collisions between electrons and neutral atoms or Ohmic electron heating within the discharge. A Druyvesteyn distribution is shown in Figure \ref{eedffig} (b) for comparison.   As time evolves, the cold electron population develops into a more Maxwellian like distribution.  This is in agreement with experimental findings for the low energy part of the EEPF, when measured with Langmuir probe. The measured EEPF has been observed to be Druyvesteyn-like early in the pulse and become more Maxwellian-like as time evolves in a discharges with tantalum \citep{gudmundsson02:249} and copper  \citep{pajdarova09:025008,sigurjonsson09:234,gudmundsson09:123302} targets.   At high electron densities electron-electron Coulomb collisions are an important energy transfer mechanism that leads to equalization of the electron distribution temperature. 
At higher electron energy, a plateau-like high energy tail appears. At  30 and 80 $\mu$s into the pulse the low energy part has developed an almost linear decrease in electron density with increased electron energy, indicating that this part is closely following a Maxwellian-like distribution. The high energy tail remains throughout the pulse.
Furthermore, secondary electrons are accelerated across the cathode  sheath and enter the IR with an energy that corresponds to the sheath voltage  $eV_{\rm SH}$ = 460 eV. These electrons show up as a peak in the electron density at around 460 eV in the EEPF calculated by OBELIX.  Note that, for a well established discharge, this peak  represents about $10^{-5}$ to $10^{-4}$ in relative density compared to 'cold' Maxwell electrons composing the discharge bulk. These high-energy  electrons cool down through collisions with neutral species and  eventually become thermalized with the colder electron population. The  most energetic electrons loose their energy largely in collisions with the  abundant argon ground state, which is why a second peak shows up at $\sim$12 eV below the energy at which the electrons are injected. 
These $\sim$12 eV correspond to the first excitation energy of the argon ground state, to the 4s levels (11.548 -- 11.828 eV). 

Besides this feature at the high-energy end of the distribution function, the EEPF calculated by OBELIX can be separated into two parts, which both show an almost linear decrease with increased electron energy, but with different slopes. 
 The comparison shows a very good agreement between the EEPF calculated from the Boltzmann equation and the two Maxwellian distributions used in the IRM. Close to the end of the pulse ($t$ = 80 $\mu$s) the IRM-assumed EEPF shows a good match with the EEPF calculated by OBELIX. At medium energies ( $\sim$ 30 -- 60 eV) the IRM-assumed  EEPF slightly overestimates the electron population, while at energies around $e V_{\rm SH}$ the high energy peak is missing in the IRM assumption.  At this electron energy the IRM-assumed EEPF  underestimates the electron density roughly by more than one  order of magnitude.

Figure \ref{ratefig} shows the temporal variation of the  electron impact excitation and ionization reaction rates 
from the IRM for hot and cold electrons and the reaction rates calculated by OBELIX.  
 All the reaction rates  involving electrons in OBELIX  are calculated from cross sections using the self-consistently determined EEDF.     Overall, the electron impact reaction rates for the cold electrons in the IRM show excellent agreement with the electron impact reaction rates  calculated by the OBELIX using the EEDF from the self-consistent calculations.  The electron impact reaction rates for excitation to the  metastable argon levels calculated by OBELIX are slightly higher than the ones used by the IRM for the cold electron group  as seen in Figure \ref{ratefig} (a).  The reaction rates for excitation to the  metastable argon levels by hot electrons are much smaller than for cold electrons.  The reaction rates for electron impact ionization from the metastable argon levels calculated by OBELIX match well with the reaction rates for the cold electron group used by the IRM, while the reaction rates for  hot electrons are much lower as seen in Figure \ref{ratefig} (b).  The reaction rate  for electron impact ionization from the argon ground state calculated by OBELIX is somewhat higher than the reaction rate from the  IRM for cold electrons, in particularly later in the pulse, as shown in Figure \ref{ratefig} (c).  The reaction rates for electron impact ionization of argon from the ground state by hot electrons used in the IRM are similar to those for the cold electrons  showing the importance of hot electrons for ionization.   The reaction rates  for electron impact ionization from the Ti ground state calculated by OBELIX show excellent agreement with the reaction rates  used by IRM for cold electrons. The reaction rates  for electron impact ionization of Ti$^+$ to create Ti$^{2+}$ calculated by OBELIX show excellent agreement for most of the duration of the pulse except towards the end where it is  slightly lower than the rate coefficients used by IRM for cold electrons as seen in Figure  \ref{ratefig} (d). Note that in all cases the hot electrons contribute much less to the overall reaction rate than the cold electrons  with the exception of electron impact ionization from ground state argon, where the ionization rates from the cold and the hot electron groups are almost equal.

\subsection{The energy cost of ionization}

\begin{figure}
\begin{center}
\includegraphics[scale=0.36]{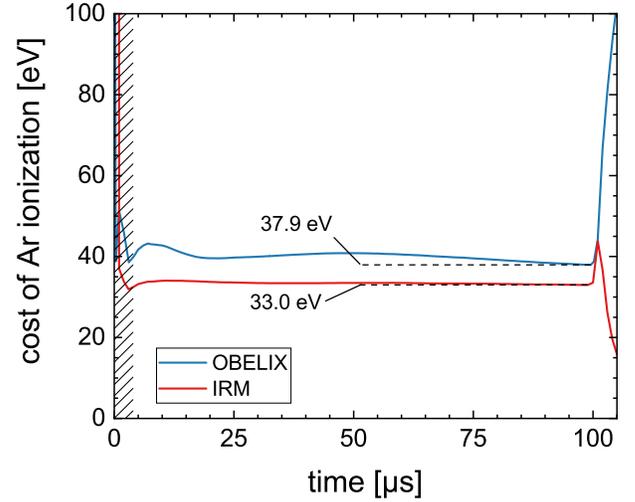}
\end{center}
 \caption{The temporal variation of the cost of ionization (COI) for the production of  Ar$^+$ ions 
during the pulse  
for the discharge with peak discharge current $I_{\rm D,peak}$ = 41 A. The blue curve shows  the results from OBELIX.   The red curve is the COI used in the IRM  properly weighted with the relative ionization reaction rates from the cold and the hot electron population.    
\label{coifig}}
\end{figure}

The concept  of collisional energy loss per electron–ion pair created, or the energy cost of ionization (COI), is often used  to calculate the ionization rate in a low temperature plasma discharge from a limited reaction set \citep{gudmundsson16:065004}. 
Since typically ionization takes place  along with electronic excitation and electron scattering, the energy cost for each ionization event is larger than simply the ionization threshold energy.  COI represents the total energy spent by energetic electrons to create an electron-ion pair. 
COI varies depending on the effective electron temperature associated with the assumed  electron energy distribution. 
Furthermore, the COI are treated in different ways in the IRM and OBELIX. 
In the IRM, the EEDF is approximated using two Maxwellian distributions, which is why the IRM includes two parts of the COI function, one for low energy electrons (electron temperature range 1 -- 7 eV) and another for high electron energies (electron temperature range 200 to 1000 eV). Furthermore, the COI used in the IRM assumes only electron impact excitation and ionization from the ground state atom.  
For comparison, the COI in  OBELIX is obtained from cross sections using the EEDF that is calculated self-consistently.
The calculation takes into account electron impact excitation and ionization from the ground state and all the excited levels of the argon atom. 
The COI per electron-ion pair created $\Epsilon_\mathrm{c}^{(X)}$, is given as \cite{lieberman05}
\begin{equation}
{\cal E}^{(X)}_{\mathrm{c}} = \frac{\sum_{{i}} {\cal E}^{(X)}_{\mathrm{iz},i} k^{(X)}_{{\mathrm{iz}},i}  
+ \sum_{{i}} \sum_j  {\cal E}^{(X)}_{{\mathrm{exc}},ji} k^{(X)}_{{\mathrm{exc}},ji} 
+ k^{(X)}_{\mathrm{el}} \frac{3m_{\mathrm{e}}}{m^{(X)}}{T_{\rm e}}}
{\sum_{{i}} k^{(X)}_{\mathrm{iz},i}} \label{eq:Ec}
\end{equation}
where $\Epsilon^{(X)}_{\mathrm{iz},i}$ is the ionization energy of species
$X$ from the ground stated and the various excited levels $i$, $\Epsilon^{(X)}_{\mathrm{exc},ji}$ and $k^{(X)}_{\mathrm{exc},ji}$ are the
excitation energy and rate coefficient for the $i$-th excitation
process of species $X$ from level $j$, respectively, $k^{(X)}_\mathrm{el}$ is the elastic scattering rate coefficient of species
$X$, $m_\mathrm{e}$ is the electron mass and $m^{(X)}$ is the mass of
species $X$. 
 Here, we compare the temporal variations of  COI  for the creation of Ar$^+$ ions used in the IRM, that was calculated prior to the run \citep{gudmundsson16:065004}  and the evolution of the COI during the pulse calculated by the  OBELIX model in Figure \ref{coifig}.  
The COI for argon from the OBELIX calculation peaks at the beginning of the pulse and then falls and  approaches 37.9 eV well into the pulse as seen in Figure \ref{coifig}. 
For comparison we show the COI used in the IRM weighted  for cold and hot electrons.  It has the value 33 eV for most of the pulse duration.
Note however that the value of the COI before 4 $\mu$s is questionableas discussed above.
 In general, the COI for Ar considered in the IRM and calculated by OBELIX match very well between the two models.  

\begin{figure}
\begin{center}
\includegraphics[scale=0.36]{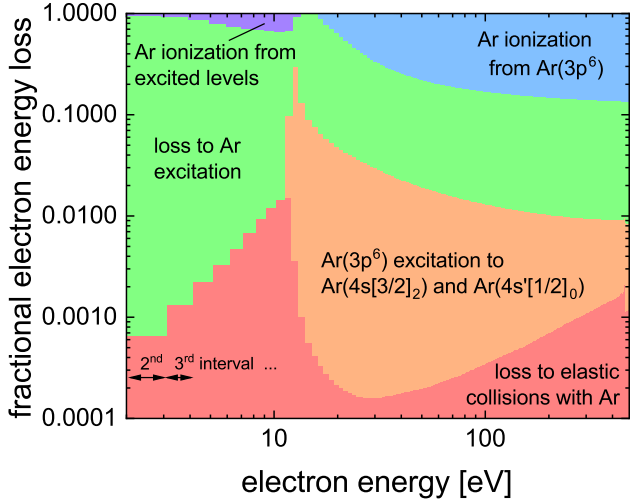}
\end{center}
 \caption{Fractional loss of electron energy by electron impact, elastic collisions, excitation and ionization as a function of electron energy calculated by OBELIX  80 $\mu$s  into the pulse. The loss processes are electron elastic collisions (red), electron impact excitation to the argon metastable states (orange), electron impact excitation to other excited argon levels (green), electron impact ionization from the higher excited levels (purple) and electron impact ionization from the ground level (blue).  To demonstrate the energy discretization used in OBELIX, the second  and third energy intervals are indicated. Note that the loss to the excitation of the argon atom is a net loss corrected for the energy gain from argon collisional de-excitation. 
\label{cofracfig}}
\end{figure}

Figure \ref{cofracfig} shows the fractional electron energy loss to the various processes, electron impact, elastic collision, electron impact excitation and electron impact ionization versus the electron energy for argon determined by the OBELIX model.  
Note that this diagram shows the electron energy loss 80 $\mu$s into the pulse. It therefore considers the loss on Ar with populated energy levels.  As a result, electrons with energies as low as few tens of meV can contribute to further excitation of the argon atom as the higher levels of this atom are closely spaced in energy.  This is different from the COI calculation used in the IRM which considers only electron impact excitation from the ground state (see e.g.~\citet[Section 3.5.]{lieberman05}). 
At 4 eV, multi-step ionization from the metastable levels starts to make a contribution, but its contribution to the  electron energy loss  in that energy interval
 is small  or up to 1.6 \% while at around 10 eV its contribution is up to 35 \%.    At 15.76 eV, ionization from the ground state sets in and its role increases with increased electron energy.  At 100 eV, it takes roughly 85 \% of the electron energy.  The COI for high energy electrons is always lower compared to the COI for low energy electrons. This is because at low electron energy, a significant portion of the energy goes to excitation and elastic collisions (Figure \ref{cofracfig}). At high electron energy, over 85 \% of the electron energy lost goes to ionization. Therefore, taking only electrons at a high energy of $> 50$ eV, the cost of ionization is very close to the actual ionization energy for the ground state argon atom. 
This is one of the reasons why the HiPIMS discharges become more efficient at higher discharge currents
\citep{brenning20:033008,brenning21:015015}. 

\begin{figure}
\begin{center}
\includegraphics[scale=0.36]{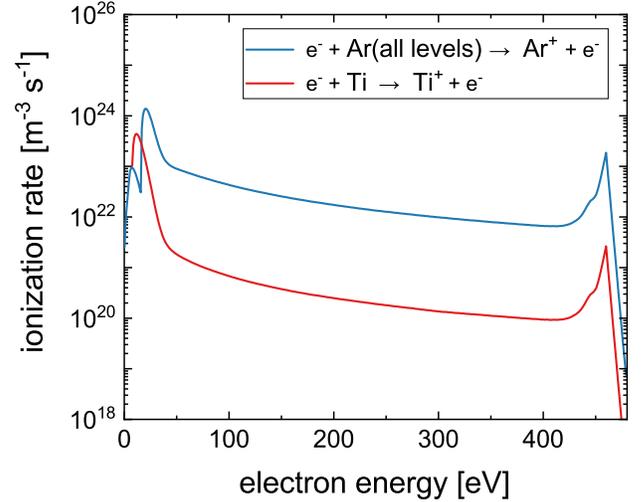}
\end{center}
 \caption{The variation of the ionization rate with electron energy  for argon and titanium calculated by  OBELIX at a time $t = 80$ $\mu$s into the pulse. 
\label{ionratefig}}
\end{figure}

A comparison of ionization rates from ground state argon and titanium is shown in Figure   \ref{ionratefig}. 
While high energy electrons are most efficient in ionizing argon, the highest ionization rate is still produced by low energy electrons. This is due to the much higher electron densities at low energies compared to high energies (Figure \ref{eedffig}). The ionization rate from all argon levels as a function of electron energy is shown in Figure \ref{ionratefig}. The ionization rate peaks at 20.9 eV for argon, which lies between the maximum of the ionization cross section and the maximum of the EEDF. A second smaller peak is situated at 7.5 eV, which lies between the maximum of the ionization cross-section from from the argon metastable levels and the maximum of the EEDF. The ionization rate falls from its peak value with increasing electron energies but peaks again at an energy corresponding to the sheath potential $eV_{\rm SH}$. For comparison, the ionization rate of titanium is shown in Figure  \ref{ionratefig}  as well. It peaks at 11.6 eV, an energy between the maximum of the ionization cross section and the maximum of the EEDF. At this energy, the ionization rate of titanium is higher compared to that of argon. 
Overall, however, the ionization rate of titanium remains well below that of argon. Just as for argon, also for titanium, a peak in ionization rate is observed at an energy corresponding to the sheath potential $eV_{\rm SH}$. Although the ionization rate has a maximum at low electron energy, the secondary electrons contribute substantially to the overall ionization rate. For argon, the ionization rate at $eV_{\rm SH}$ is around 10 \% of the maximum ionization rate at the low energy peak. For Ti, this is around 5 \% of the maximum ionization rate at  the low energy peak. This substantial contribution is despite the much lower electron density at this high energy (Figure \ref{eedffig}) and is  an effect of the more efficient use of high energy electrons compared to low energy electrons for ionization.

\section{Conclusion}
\label{conclusion}

We have applied  the ionization region model (IRM) and the OBELIX model to study the electron kinetics in a HiPIMS discharge.  The two models are merged in the sense that the results of the IRM is used as an input for OBELIX. 
 In the IRM, the bulk (cold)  electrons are assumed to exhibit a Maxwellian distribution and the secondary electrons are taken as a high-energy tail, while in OBELIX, the electron energy distribution  is calculated self-consistently using an isotropic Boltzmann equation.
There is generally a very good agreement between the two models for the temporal evolution of the particle densities. 
Furthermore, there is a very good agreement between the bi-Maxwellian electron energy distribution assumed by the IRM and the  electron energy distribution that is calculated self-consistently using the OBELIX model.  Therefore, assuming a bi-Maxwellian EEDF that constitutes  cold bulk electron  and hot secondary electron groups appears to be a good approximation when modeling the HiPIMS discharge. 
Although this assumption was well justified \citep{huo13:045005,huo17:354003}, a comparison with a self-consistent solution of the Boltzmann equation  has not been shown to date.
These results can be taken as an additional step for the validation of the IRM approach to model HiPIMS discharges.    
The purpose of the IRM, with a minimum computational time, is to reproduce internal HiPIMS discharge parameters accurately enough to give insight into the physics. In the studies reported here, an IRM run takes approximately 0.02 \% 
of the time for an OBELIX run at the cost of only a minor loss in precision of the discharge kinetics. 
This means that the IRM assumptions of a bi-Maxwellian electron population, and of a simplified excitation schemes, are justified.  

\appendix
\section{Comparison of cross sections}
\label{crosssectionappendix}

\begin{figure*}
\begin{center}
\includegraphics[scale=0.32]{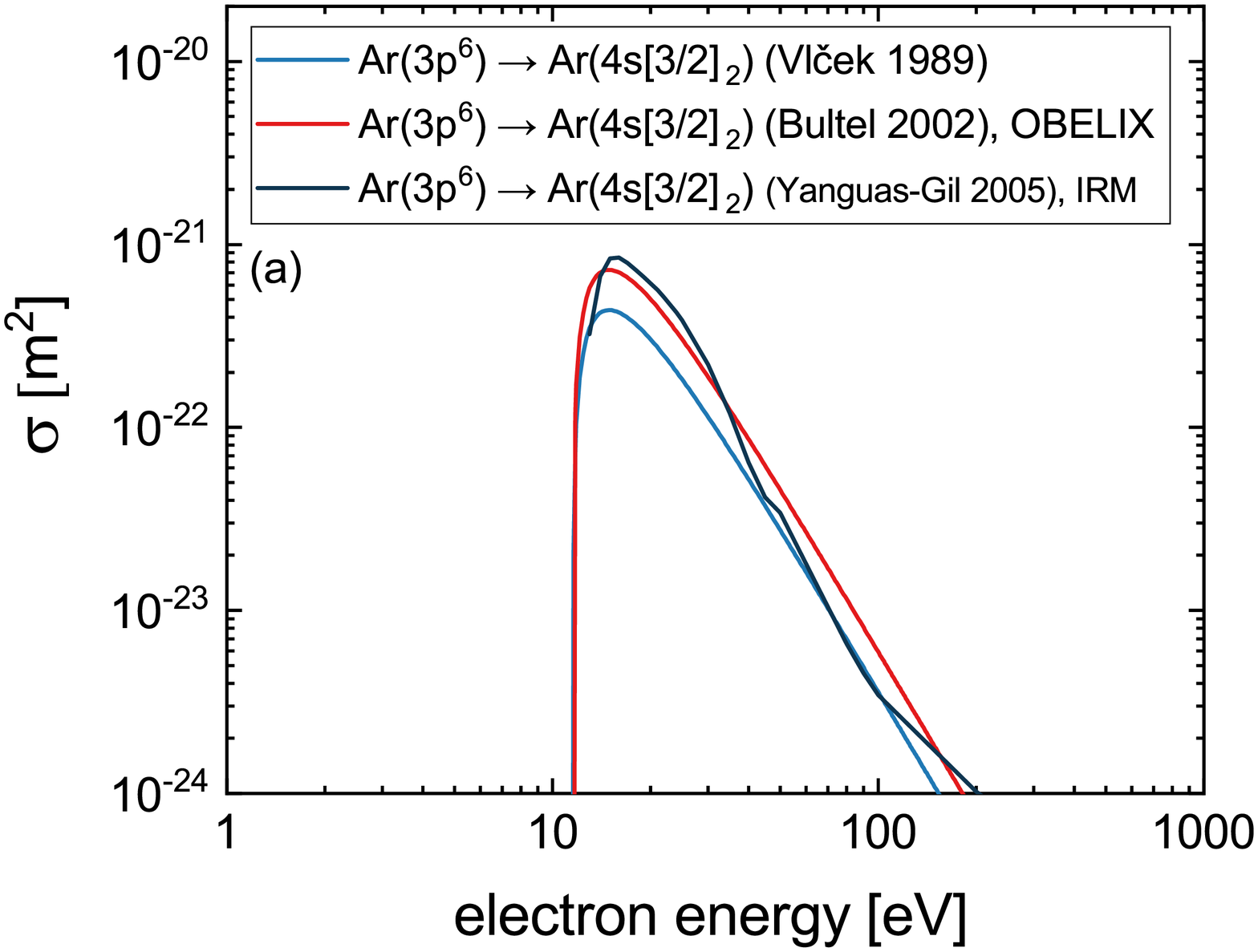}
\includegraphics[scale=0.32]{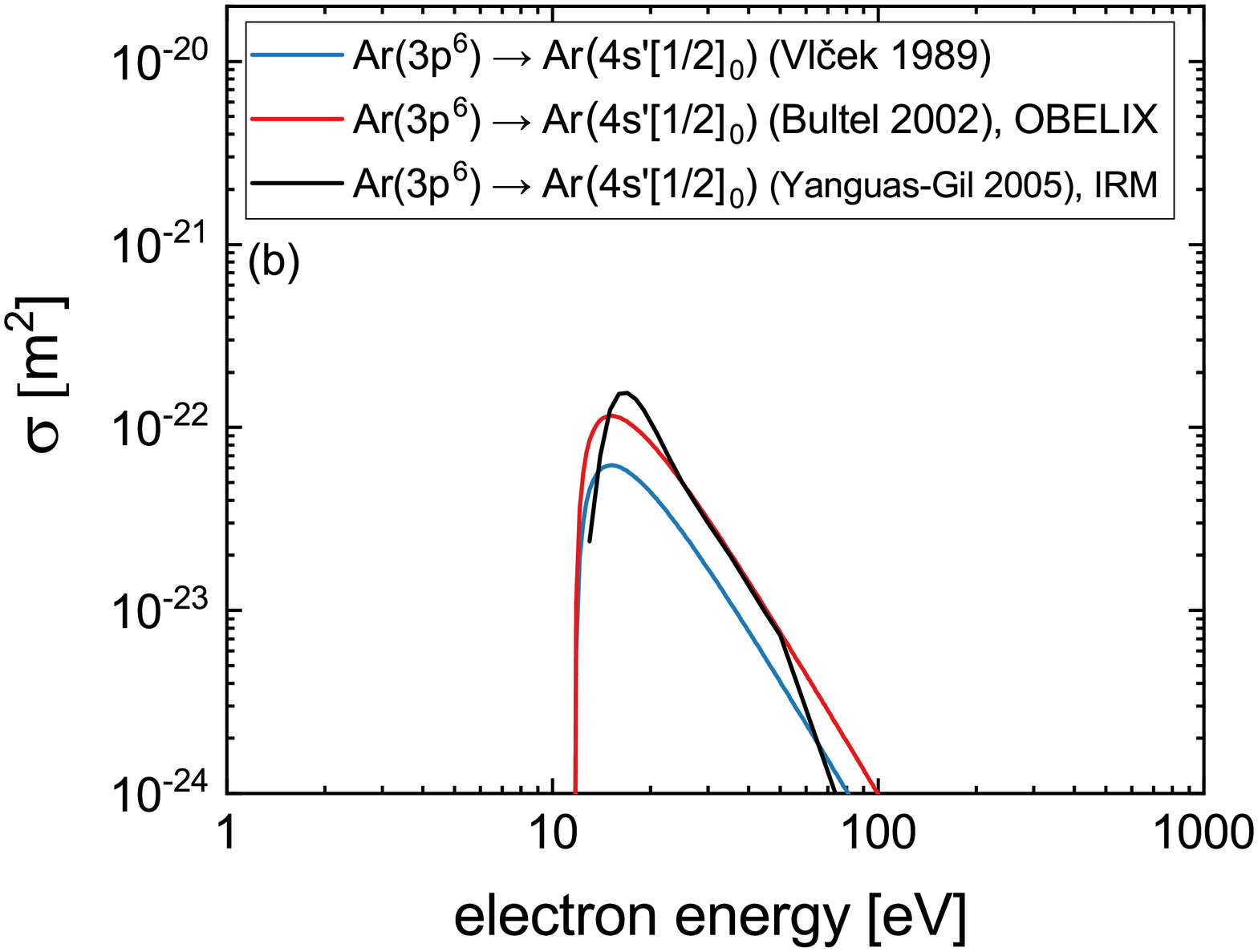}
\includegraphics[scale=0.32]{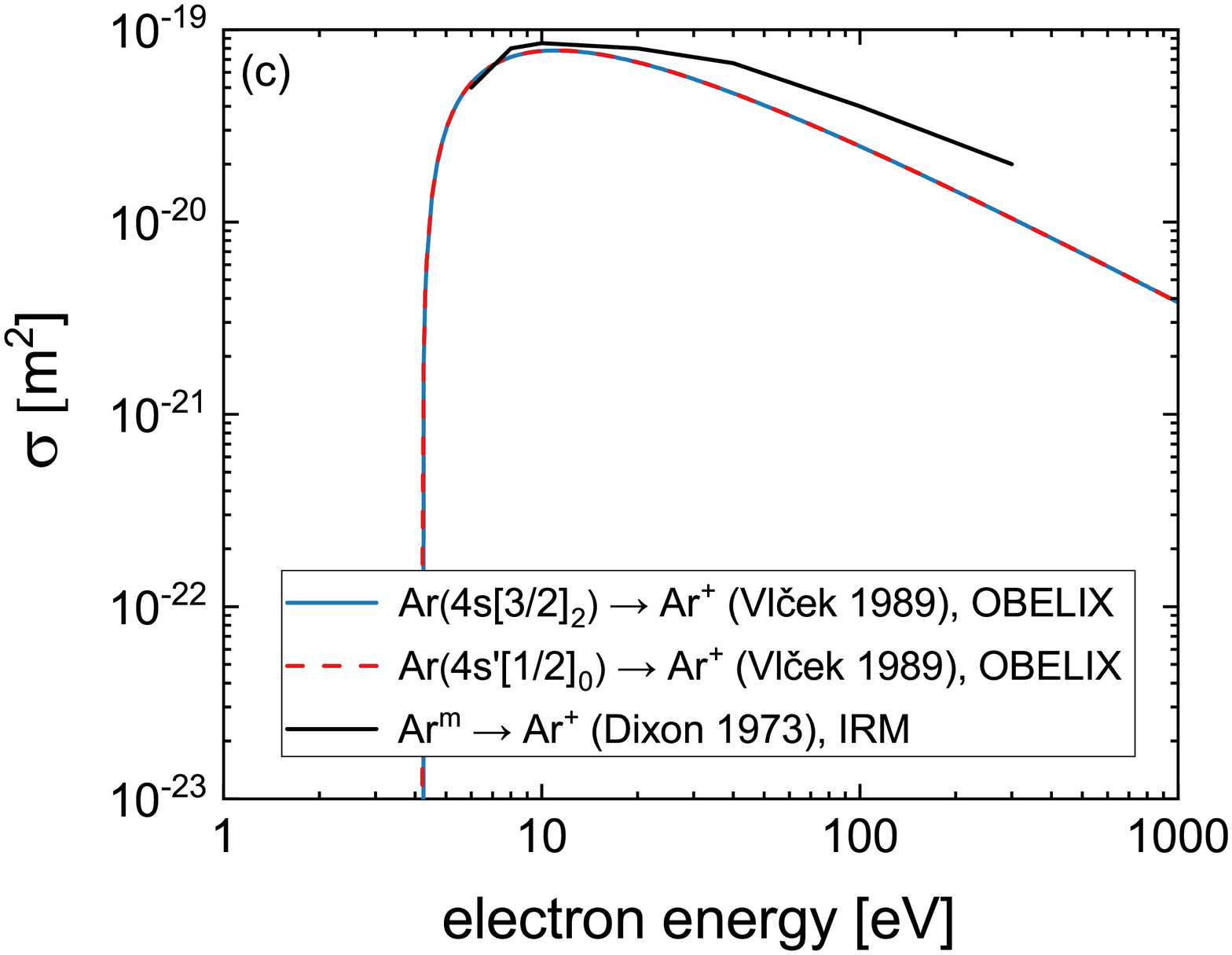}
\includegraphics[scale=0.32]{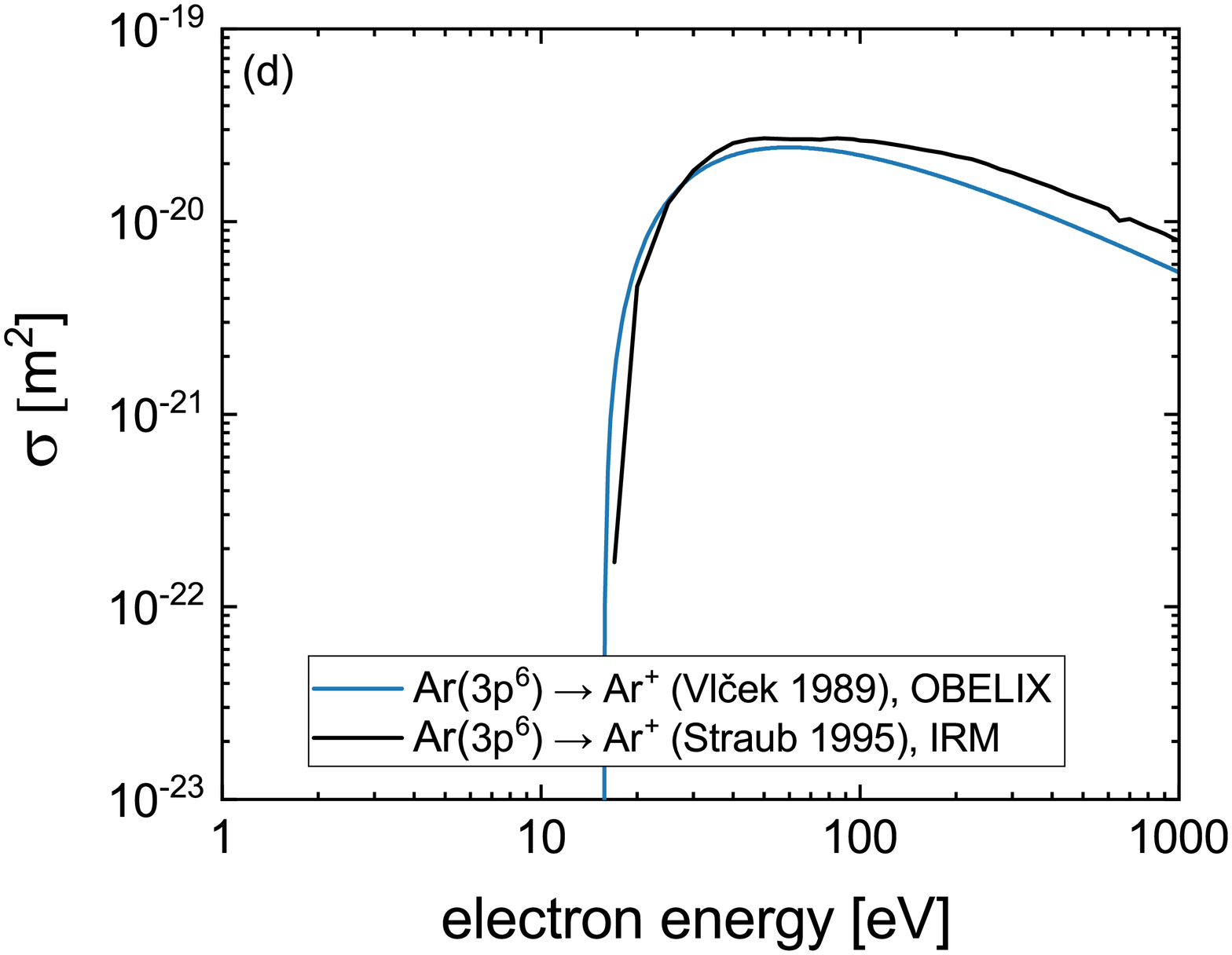}
\end{center}
 \caption{Comparison of electron impact excitation cross sections used in OBELIX and IRM  from the ground state Ar(3p$^6$) (a) to the metastable level Ar(4s[3/2]$_2$) and (b) to the metastable level Ar(4s'[1/2]$_0$) and a comparison of the electron impact ionization from the (c) metastable levels 
 Ar(4s[3/2]$_2$)  and  Ar(4s'[1/2]$_0$), and (d)  the ground state Ar(3p$^6$). Note
that the combined metastable levels (Ar(4s'[1/2]$_0$) + Ar(4s[3/2]$_2$)) are denoted by Ar$^{\rm m}$.
\label{crosse}}
\end{figure*}

The cross sections used in the IRM and OBELIX are taken from different sources. 
For the IRM, the rate coefficients for the cold and hot electrons are calculated using the chosen cross section assuming a Maxwellian EEDF as discussed in Section \ref{reactionsetsIRM}. 
  Here we compare  electron impact excitation and ionization cross sections used for reactions in both models.  
 The cross sections used  in OBELIX and IRM for electron impact excitation from the ground state Ar(3p$^6$) to the metastable level Ar(4s[3/2]$_2$) are shown in
Figure \ref{crosse} (a) and to the metastable level Ar(4s'[1/2]$_0$) in Figure \ref{crosse} (b).  Figures \ref{crosse} (a)  and (b) show that Drawin’s empirical electron impact excitation cross sections using the most recent fitting parameters    \citep{bultel02:046406}  correspond well to the cross sections given in a review by  \citet{yanguasgil05:1588}, originating from  \citet{khakoo04:247}.  Figure \ref{crosse} (c) shows the  electron impact ionization cross sections  from the  metastable levels 
 Ar(4s[3/2]$_2$) and  Ar(4s'[1/2]$_0$) and Figure \ref{crosse} (d) shows the electron impact ionization cross section from  the ground state Ar(3p$^6$). 
Figures \ref{crosse} (c) and (d) show a good match between the ionization cross sections from ground-state and from each of the two metastable states (Ar(4s[3/2]$_2$) and Ar(4s'[1/2]$_0$)) to experimentally determined cross sections by  \citet{dixon73:405} as well as from the ground state experimentally determined by \citet{straub95:1115}. 

\acknowledgments

M.~R.~is grateful to Dr.~Dmitry Kalanov at the Leibniz Institute of Surface Engineering (IOM),  Prof.~Annemie Bogaerts of the University of Antwerp  and Prof.~Jaroslav Vl\v{c}ek of the University of West Bohemia for very fruitful discussions and advice on electron impact excitations of argon and the argon model.  
This work was partially supported by the Free State of Saxony and the European Regional Development Fund (Grant No.~100336119),
the  Icelandic Research Fund (Grant No.~196141), the Swedish Research Council (Grant VR 2018-04139) and the Swedish Government Strategic Research Area in Materials Science on Functional Materials at Link{\"o}ping University (Faculty Grant SFO-Mat-LiU No.~2009-00971).  

\vspace{3 cm}

Martin Rudolph 

{\tt https://orcid.org/0000-0002-0854-6708}

Nils Brenning

{\tt https://orcid.org/0000-0003-1308-9270}

Adrien Revel 

{\tt https://orcid.org/0000-0003-2758-2058}

Hamidreza Hajihoseini 

{\tt https://orcid.org/0000-0002-2494-6584}

Jon Tomas Gudmundsson 

{\tt https://orcid.org/0000-0002-8153-3209}

Andr{\'e} Anders 

{\tt https://orcid.org/0000-0002-5313-6505}

Tiberiu M. Minea

{\tt  0000-0003-2886-3492 }

Daniel Lundin 

{\tt https://orcid.org/0000-0001-8591-1003}


\begin{thebibliography}{86}
\expandafter\ifx\csname natexlab\endcsname\relax\def\natexlab#1{#1}\fi
\expandafter\ifx\csname bibnamefont\endcsname\relax
  \def\bibnamefont#1{#1}\fi
\expandafter\ifx\csname bibfnamefont\endcsname\relax
  \def\bibfnamefont#1{#1}\fi
\expandafter\ifx\csname citenamefont\endcsname\relax
  \def\citenamefont#1{#1}\fi
\expandafter\ifx\csname url\endcsname\relax
  \def\url#1{\texttt{#1}}\fi
\expandafter\ifx\csname urlprefix\endcsname\relax\def\urlprefix{URL }\fi
\providecommand{\bibinfo}[2]{#2}
\providecommand{\eprint}[2][]{\url{#2}}

\bibitem[{\citenamefont{Gudmundsson}(2020)}]{gudmundsson20:113001}
\bibinfo{author}{\bibfnamefont{J.~T.} \bibnamefont{Gudmundsson}},
  \bibinfo{journal}{Plasma Sources Science and Technology}
  \textbf{\bibinfo{volume}{29}}, \bibinfo{pages}{113001}
  (\bibinfo{year}{2020}).

\bibitem[{\citenamefont{Gudmundsson et~al.}(2012)\citenamefont{Gudmundsson,
  Brenning, Lundin, and Helmersson}}]{gudmundsson12:030801}
\bibinfo{author}{\bibfnamefont{J.~T.} \bibnamefont{Gudmundsson}},
  \bibinfo{author}{\bibfnamefont{N.}~\bibnamefont{Brenning}},
  \bibinfo{author}{\bibfnamefont{D.}~\bibnamefont{Lundin}}, \bibnamefont{and}
  \bibinfo{author}{\bibfnamefont{U.}~\bibnamefont{Helmersson}},
  \bibinfo{journal}{Journal of Vacuum Science and Technology A}
  \textbf{\bibinfo{volume}{30}}, \bibinfo{pages}{030801}
  (\bibinfo{year}{2012}).

\bibitem[{\citenamefont{Hubi\v{c}ka et~al.}(2020)\citenamefont{Hubi\v{c}ka,
  Gudmundsson, Larsson, and Lundin}}]{hubicka20:49}
\bibinfo{author}{\bibfnamefont{Z.}~\bibnamefont{Hubi\v{c}ka}},
  \bibinfo{author}{\bibfnamefont{J.~T.} \bibnamefont{Gudmundsson}},
  \bibinfo{author}{\bibfnamefont{P.}~\bibnamefont{Larsson}}, \bibnamefont{and}
  \bibinfo{author}{\bibfnamefont{D.}~\bibnamefont{Lundin}}, in
  \emph{\bibinfo{booktitle}{High Power Impulse Magnetron Sputtering:
  {F}undamentals, Technologies, Challenges and Applications}}, edited by
  \bibinfo{editor}{\bibfnamefont{D.}~\bibnamefont{Lundin}},
  \bibinfo{editor}{\bibfnamefont{T.}~\bibnamefont{Minea}}, \bibnamefont{and}
  \bibinfo{editor}{\bibfnamefont{J.~T.} \bibnamefont{Gudmundsson}}
  (\bibinfo{publisher}{Elsevier}, \bibinfo{address}{Amsterdam, The
  Netherlands}, \bibinfo{year}{2020}), pp. \bibinfo{pages}{49--80}.

\bibitem[{\citenamefont{Anders}(2011)}]{anders11:S1}
\bibinfo{author}{\bibfnamefont{A.}~\bibnamefont{Anders}},
  \bibinfo{journal}{Surface and Coatings Technology}
  \textbf{\bibinfo{volume}{205}}, \bibinfo{pages}{S1} (\bibinfo{year}{2011}).

\bibitem[{\citenamefont{Lundin and Sarakinos}(2012)}]{lundin12:780}
\bibinfo{author}{\bibfnamefont{D.}~\bibnamefont{Lundin}} \bibnamefont{and}
  \bibinfo{author}{\bibfnamefont{K.}~\bibnamefont{Sarakinos}},
  \bibinfo{journal}{Journal of Materials Research}
  \textbf{\bibinfo{volume}{27}}, \bibinfo{pages}{780} (\bibinfo{year}{2012}).

\bibitem[{\citenamefont{Greczynski et~al.}(2019)\citenamefont{Greczynski,
  Petrov, Greene, and Hultman}}]{greczynski19:060801}
\bibinfo{author}{\bibfnamefont{G.}~\bibnamefont{Greczynski}},
  \bibinfo{author}{\bibfnamefont{I.}~\bibnamefont{Petrov}},
  \bibinfo{author}{\bibfnamefont{J.~E.} \bibnamefont{Greene}},
  \bibnamefont{and} \bibinfo{author}{\bibfnamefont{L.}~\bibnamefont{Hultman}},
  \bibinfo{journal}{Journal of Vacuum Science and Technology A}
  \textbf{\bibinfo{volume}{37}}, \bibinfo{pages}{060801}
  (\bibinfo{year}{2019}).

\bibitem[{\citenamefont{Greczynski et~al.}(2020)\citenamefont{Greczynski,
  Mr{\'a}z, Schneider, and Hultman}}]{greczynski20:180901}
\bibinfo{author}{\bibfnamefont{G.}~\bibnamefont{Greczynski}},
  \bibinfo{author}{\bibfnamefont{S.}~\bibnamefont{Mr{\'a}z}},
  \bibinfo{author}{\bibfnamefont{J.~M.} \bibnamefont{Schneider}},
  \bibnamefont{and} \bibinfo{author}{\bibfnamefont{L.}~\bibnamefont{Hultman}},
  \bibinfo{journal}{Journal of Applied Physics} \textbf{\bibinfo{volume}{127}},
  \bibinfo{pages}{180901} (\bibinfo{year}{2020}).

\bibitem[{\citenamefont{Petrov et~al.}(1993)\citenamefont{Petrov, Adibi,
  Greene, Hultman, and Sundgren}}]{petrov93:36}
\bibinfo{author}{\bibfnamefont{I.}~\bibnamefont{Petrov}},
  \bibinfo{author}{\bibfnamefont{F.}~\bibnamefont{Adibi}},
  \bibinfo{author}{\bibfnamefont{J.~E.} \bibnamefont{Greene}},
  \bibinfo{author}{\bibfnamefont{L.}~\bibnamefont{Hultman}}, \bibnamefont{and}
  \bibinfo{author}{\bibfnamefont{J.-E.} \bibnamefont{Sundgren}},
  \bibinfo{journal}{Applied Physics Letters} \textbf{\bibinfo{volume}{63}},
  \bibinfo{pages}{36} (\bibinfo{year}{1993}).

\bibitem[{\citenamefont{Petrov et~al.}(2003)\citenamefont{Petrov, Barna,
  Hultman, and Greene}}]{petrov03:S117}
\bibinfo{author}{\bibfnamefont{I.}~\bibnamefont{Petrov}},
  \bibinfo{author}{\bibfnamefont{P.~B.} \bibnamefont{Barna}},
  \bibinfo{author}{\bibfnamefont{L.}~\bibnamefont{Hultman}}, \bibnamefont{and}
  \bibinfo{author}{\bibfnamefont{J.~E.} \bibnamefont{Greene}},
  \bibinfo{journal}{Journal of Vacuum Science and Technology A}
  \textbf{\bibinfo{volume}{21}}, \bibinfo{pages}{S117} (\bibinfo{year}{2003}).

\bibitem[{\citenamefont{Alami et~al.}(2005)\citenamefont{Alami, Petersson,
  Music, Gudmundsson, Bohlmark, and Helmersson}}]{alami05:278}
\bibinfo{author}{\bibfnamefont{J.}~\bibnamefont{Alami}},
  \bibinfo{author}{\bibfnamefont{P.~O.~A.} \bibnamefont{Petersson}},
  \bibinfo{author}{\bibfnamefont{D.}~\bibnamefont{Music}},
  \bibinfo{author}{\bibfnamefont{J.~T.} \bibnamefont{Gudmundsson}},
  \bibinfo{author}{\bibfnamefont{J.}~\bibnamefont{Bohlmark}}, \bibnamefont{and}
  \bibinfo{author}{\bibfnamefont{U.}~\bibnamefont{Helmersson}},
  \bibinfo{journal}{Journal of Vacuum Science and Technology A}
  \textbf{\bibinfo{volume}{23}}, \bibinfo{pages}{278} (\bibinfo{year}{2005}).

\bibitem[{\citenamefont{Sarakinos and Martinu}(2020)}]{sarakinos20:333}
\bibinfo{author}{\bibfnamefont{K.}~\bibnamefont{Sarakinos}} \bibnamefont{and}
  \bibinfo{author}{\bibfnamefont{L.}~\bibnamefont{Martinu}}, in
  \emph{\bibinfo{booktitle}{High Power Impulse Mangetron Sputtering:
  {F}undamentals, Technologies, Challenges and Applications}}, edited by
  \bibinfo{editor}{\bibfnamefont{D.}~\bibnamefont{Lundin}},
  \bibinfo{editor}{\bibfnamefont{T.}~\bibnamefont{Minea}}, \bibnamefont{and}
  \bibinfo{editor}{\bibfnamefont{J.~T.} \bibnamefont{Gudmundsson}}
  (\bibinfo{publisher}{Elsevier}, \bibinfo{address}{Amsterdam, The
  Netherlands}, \bibinfo{year}{2020}), pp. \bibinfo{pages}{333--374}.

\bibitem[{\citenamefont{Cemin et~al.}(2017)\citenamefont{Cemin, Abadias, Minea,
  Furgeaud, Brisset, Solas, and Lundin}}]{cemin17:120}
\bibinfo{author}{\bibfnamefont{F.}~\bibnamefont{Cemin}},
  \bibinfo{author}{\bibfnamefont{G.}~\bibnamefont{Abadias}},
  \bibinfo{author}{\bibfnamefont{T.}~\bibnamefont{Minea}},
  \bibinfo{author}{\bibfnamefont{C.}~\bibnamefont{Furgeaud}},
  \bibinfo{author}{\bibfnamefont{F.}~\bibnamefont{Brisset}},
  \bibinfo{author}{\bibfnamefont{D.}~\bibnamefont{Solas}}, \bibnamefont{and}
  \bibinfo{author}{\bibfnamefont{D.}~\bibnamefont{Lundin}},
  \bibinfo{journal}{Acta Materialia} \textbf{\bibinfo{volume}{141}},
  \bibinfo{pages}{120} (\bibinfo{year}{2017}).

\bibitem[{\citenamefont{Hajihoseini et~al.}(2018)\citenamefont{Hajihoseini,
  Kateb, Ingvarsson, and Gudmundsson}}]{hajihoseini18:126}
\bibinfo{author}{\bibfnamefont{H.}~\bibnamefont{Hajihoseini}},
  \bibinfo{author}{\bibfnamefont{M.}~\bibnamefont{Kateb}},
  \bibinfo{author}{\bibfnamefont{S.}~\bibnamefont{Ingvarsson}},
  \bibnamefont{and} \bibinfo{author}{\bibfnamefont{J.~T.}
  \bibnamefont{Gudmundsson}}, \bibinfo{journal}{Thin Solid Films}
  \textbf{\bibinfo{volume}{663}}, \bibinfo{pages}{126} (\bibinfo{year}{2018}).

\bibitem[{\citenamefont{Sheridan et~al.}(1991)\citenamefont{Sheridan, Goeckner,
  and Goree}}]{sheridan91:688}
\bibinfo{author}{\bibfnamefont{T.~E.} \bibnamefont{Sheridan}},
  \bibinfo{author}{\bibfnamefont{M.~J.} \bibnamefont{Goeckner}},
  \bibnamefont{and} \bibinfo{author}{\bibfnamefont{J.}~\bibnamefont{Goree}},
  \bibinfo{journal}{Journal of Vacuum Science and Technology A}
  \textbf{\bibinfo{volume}{9}}, \bibinfo{pages}{688} (\bibinfo{year}{1991}).

\bibitem[{\citenamefont{Seo et~al.}(2004)\citenamefont{Seo, In, and
  Chang}}]{seo04:409}
\bibinfo{author}{\bibfnamefont{S.-H.} \bibnamefont{Seo}},
  \bibinfo{author}{\bibfnamefont{J.-H.} \bibnamefont{In}}, \bibnamefont{and}
  \bibinfo{author}{\bibfnamefont{H.-Y.} \bibnamefont{Chang}},
  \bibinfo{journal}{Plasma Sources Science and Technology}
  \textbf{\bibinfo{volume}{13}}, \bibinfo{pages}{409} (\bibinfo{year}{2004}).

\bibitem[{\citenamefont{Gudmundsson et~al.}(2002)\citenamefont{Gudmundsson,
  Alami, and Helmersson}}]{gudmundsson02:249}
\bibinfo{author}{\bibfnamefont{J.~T.} \bibnamefont{Gudmundsson}},
  \bibinfo{author}{\bibfnamefont{J.}~\bibnamefont{Alami}}, \bibnamefont{and}
  \bibinfo{author}{\bibfnamefont{U.}~\bibnamefont{Helmersson}},
  \bibinfo{journal}{Surface and Coatings Technology}
  \textbf{\bibinfo{volume}{161}}, \bibinfo{pages}{249} (\bibinfo{year}{2002}).

\bibitem[{\citenamefont{Pajdarov{\'a} et~al.}(2009)\citenamefont{Pajdarov{\'a},
  Vl\v{c}ek, Kudl{\'a}\v{c}ek, and Luk{\'a}\v{s}}}]{pajdarova09:025008}
\bibinfo{author}{\bibfnamefont{A.~D.} \bibnamefont{Pajdarov{\'a}}},
  \bibinfo{author}{\bibfnamefont{J.}~\bibnamefont{Vl\v{c}ek}},
  \bibinfo{author}{\bibfnamefont{P.}~\bibnamefont{Kudl{\'a}\v{c}ek}},
  \bibnamefont{and}
  \bibinfo{author}{\bibfnamefont{J.}~\bibnamefont{Luk{\'a}\v{s}}},
  \bibinfo{journal}{Plasma Sources Science and Technology}
  \textbf{\bibinfo{volume}{18}}, \bibinfo{pages}{025008}
  (\bibinfo{year}{2009}).

\bibitem[{\citenamefont{Brenning et~al.}(2017)\citenamefont{Brenning,
  Gudmundsson, Raadu, Petty, Minea, and Lundin}}]{brenning17:125003}
\bibinfo{author}{\bibfnamefont{N.}~\bibnamefont{Brenning}},
  \bibinfo{author}{\bibfnamefont{J.~T.} \bibnamefont{Gudmundsson}},
  \bibinfo{author}{\bibfnamefont{M.~A.} \bibnamefont{Raadu}},
  \bibinfo{author}{\bibfnamefont{T.~J.} \bibnamefont{Petty}},
  \bibinfo{author}{\bibfnamefont{T.}~\bibnamefont{Minea}}, \bibnamefont{and}
  \bibinfo{author}{\bibfnamefont{D.}~\bibnamefont{Lundin}},
  \bibinfo{journal}{Plasma Sources Science and Technology}
  \textbf{\bibinfo{volume}{26}}, \bibinfo{pages}{125003}
  (\bibinfo{year}{2017}).

\bibitem[{\citenamefont{Minea et~al.}(2020)\citenamefont{Minea, Koz{\'a}k,
  Costin, Gudmundsson, and Lundin}}]{minea20:159}
\bibinfo{author}{\bibfnamefont{T.}~\bibnamefont{Minea}},
  \bibinfo{author}{\bibfnamefont{T.}~\bibnamefont{Koz{\'a}k}},
  \bibinfo{author}{\bibfnamefont{C.}~\bibnamefont{Costin}},
  \bibinfo{author}{\bibfnamefont{J.~T.} \bibnamefont{Gudmundsson}},
  \bibnamefont{and} \bibinfo{author}{\bibfnamefont{D.}~\bibnamefont{Lundin}},
  in \emph{\bibinfo{booktitle}{High Power Impulse Mangetron Sputtering:
  {F}undamentals, Technologies, Challenges and Applications}}, edited by
  \bibinfo{editor}{\bibfnamefont{D.}~\bibnamefont{Lundin}},
  \bibinfo{editor}{\bibfnamefont{T.}~\bibnamefont{Minea}}, \bibnamefont{and}
  \bibinfo{editor}{\bibfnamefont{J.~T.} \bibnamefont{Gudmundsson}}
  (\bibinfo{publisher}{Elsevier}, \bibinfo{address}{Amsterdam, The
  Netherlands}, \bibinfo{year}{2020}), pp. \bibinfo{pages}{159--221}.

\bibitem[{\citenamefont{Christie}(2005)}]{christie05:330}
\bibinfo{author}{\bibfnamefont{D.~J.} \bibnamefont{Christie}},
  \bibinfo{journal}{Journal of Vacuum Science and Technology A}
  \textbf{\bibinfo{volume}{23}}, \bibinfo{pages}{330} (\bibinfo{year}{2005}).

\bibitem[{\citenamefont{Vl\v{c}ek et~al.}(2007)\citenamefont{Vl\v{c}ek,
  Kudl\'{a}\v{c}ek, Burcalov\'{a}, and Musil}}]{vlcek07:42}
\bibinfo{author}{\bibfnamefont{J.}~\bibnamefont{Vl\v{c}ek}},
  \bibinfo{author}{\bibfnamefont{P.}~\bibnamefont{Kudl\'{a}\v{c}ek}},
  \bibinfo{author}{\bibfnamefont{K.}~\bibnamefont{Burcalov\'{a}}},
  \bibnamefont{and} \bibinfo{author}{\bibfnamefont{J.}~\bibnamefont{Musil}},
  \bibinfo{journal}{Journal of Vacuum Science and Technology A}
  \textbf{\bibinfo{volume}{25}}, \bibinfo{pages}{42} (\bibinfo{year}{2007}).

\bibitem[{\citenamefont{Vl\v{c}ek and Burcalov{\'a}}(2010)}]{vlcek10:065010}
\bibinfo{author}{\bibfnamefont{J.}~\bibnamefont{Vl\v{c}ek}} \bibnamefont{and}
  \bibinfo{author}{\bibfnamefont{K.}~\bibnamefont{Burcalov{\'a}}},
  \bibinfo{journal}{Plasma Sources Science and Technology}
  \textbf{\bibinfo{volume}{19}}, \bibinfo{pages}{065010}
  (\bibinfo{year}{2010}).

\bibitem[{\citenamefont{Raadu et~al.}(2011)\citenamefont{Raadu, Axn{\"a}s,
  Gudmundsson, Huo, and Brenning}}]{raadu11:065007}
\bibinfo{author}{\bibfnamefont{M.~A.} \bibnamefont{Raadu}},
  \bibinfo{author}{\bibfnamefont{I.}~\bibnamefont{Axn{\"a}s}},
  \bibinfo{author}{\bibfnamefont{J.~T.} \bibnamefont{Gudmundsson}},
  \bibinfo{author}{\bibfnamefont{C.}~\bibnamefont{Huo}}, \bibnamefont{and}
  \bibinfo{author}{\bibfnamefont{N.}~\bibnamefont{Brenning}},
  \bibinfo{journal}{Plasma Sources Science and Technology}
  \textbf{\bibinfo{volume}{20}}, \bibinfo{pages}{065007}
  (\bibinfo{year}{2011}).

\bibitem[{\citenamefont{Huo et~al.}(2017)\citenamefont{Huo, Lundin,
  Gudmundsson, Raadu, Bradley, and Brenning}}]{huo17:354003}
\bibinfo{author}{\bibfnamefont{C.}~\bibnamefont{Huo}},
  \bibinfo{author}{\bibfnamefont{D.}~\bibnamefont{Lundin}},
  \bibinfo{author}{\bibfnamefont{J.~T.} \bibnamefont{Gudmundsson}},
  \bibinfo{author}{\bibfnamefont{M.~A.} \bibnamefont{Raadu}},
  \bibinfo{author}{\bibfnamefont{J.~W.} \bibnamefont{Bradley}},
  \bibnamefont{and} \bibinfo{author}{\bibfnamefont{N.}~\bibnamefont{Brenning}},
  \bibinfo{journal}{Journal of Physics D: Applied Physics}
  \textbf{\bibinfo{volume}{50}}, \bibinfo{pages}{354003}
  (\bibinfo{year}{2017}).

\bibitem[{\citenamefont{Huo et~al.}(2013)\citenamefont{Huo, Lundin, Raadu,
  Anders, Gudmundsson, and Brenning}}]{huo13:045005}
\bibinfo{author}{\bibfnamefont{C.}~\bibnamefont{Huo}},
  \bibinfo{author}{\bibfnamefont{D.}~\bibnamefont{Lundin}},
  \bibinfo{author}{\bibfnamefont{M.~A.} \bibnamefont{Raadu}},
  \bibinfo{author}{\bibfnamefont{A.}~\bibnamefont{Anders}},
  \bibinfo{author}{\bibfnamefont{J.~T.} \bibnamefont{Gudmundsson}},
  \bibnamefont{and} \bibinfo{author}{\bibfnamefont{N.}~\bibnamefont{Brenning}},
  \bibinfo{journal}{Plasma Sources Science and Technology}
  \textbf{\bibinfo{volume}{22}}, \bibinfo{pages}{045005}
  (\bibinfo{year}{2013}).

\bibitem[{\citenamefont{Brenning et~al.}(2016)\citenamefont{Brenning,
  Gudmundsson, Lundin, Minea, Raadu, and Helmersson}}]{brenning16:065024}
\bibinfo{author}{\bibfnamefont{N.}~\bibnamefont{Brenning}},
  \bibinfo{author}{\bibfnamefont{J.~T.} \bibnamefont{Gudmundsson}},
  \bibinfo{author}{\bibfnamefont{D.}~\bibnamefont{Lundin}},
  \bibinfo{author}{\bibfnamefont{T.}~\bibnamefont{Minea}},
  \bibinfo{author}{\bibfnamefont{M.~A.} \bibnamefont{Raadu}}, \bibnamefont{and}
  \bibinfo{author}{\bibfnamefont{U.}~\bibnamefont{Helmersson}},
  \bibinfo{journal}{Plasma Sources Science and Technology}
  \textbf{\bibinfo{volume}{25}}, \bibinfo{pages}{065024}
  (\bibinfo{year}{2016}).

\bibitem[{\citenamefont{Depla et~al.}(2009)\citenamefont{Depla, Mahieu, and {De
  Gryse}}}]{depla09:2825}
\bibinfo{author}{\bibfnamefont{D.}~\bibnamefont{Depla}},
  \bibinfo{author}{\bibfnamefont{S.}~\bibnamefont{Mahieu}}, \bibnamefont{and}
  \bibinfo{author}{\bibfnamefont{R.}~\bibnamefont{{De Gryse}}},
  \bibinfo{journal}{Thin Solid Films} \textbf{\bibinfo{volume}{517}},
  \bibinfo{pages}{2825} (\bibinfo{year}{2009}).

\bibitem[{\citenamefont{Thornton}(1978)}]{thornton78:171}
\bibinfo{author}{\bibfnamefont{J.~A.} \bibnamefont{Thornton}},
  \bibinfo{journal}{Journal of Vacuum Science and Technology}
  \textbf{\bibinfo{volume}{15}}, \bibinfo{pages}{171} (\bibinfo{year}{1978}).

\bibitem[{\citenamefont{Gudmundsson and
  Hecimovic}(2017)}]{gudmundsson17:123001}
\bibinfo{author}{\bibfnamefont{J.~T.} \bibnamefont{Gudmundsson}}
  \bibnamefont{and}
  \bibinfo{author}{\bibfnamefont{A.}~\bibnamefont{Hecimovic}},
  \bibinfo{journal}{Plasma Sources Science and Technology}
  \textbf{\bibinfo{volume}{26}}, \bibinfo{pages}{123001}
  (\bibinfo{year}{2017}).

\bibitem[{\citenamefont{Panjan and Anders}(2017)}]{panjan17:063302}
\bibinfo{author}{\bibfnamefont{M.}~\bibnamefont{Panjan}} \bibnamefont{and}
  \bibinfo{author}{\bibfnamefont{A.}~\bibnamefont{Anders}},
  \bibinfo{journal}{Journal of Applied Physics} \textbf{\bibinfo{volume}{121}},
  \bibinfo{pages}{063302} (\bibinfo{year}{2017}).

\bibitem[{\citenamefont{Rudolph et~al.}(2021)\citenamefont{Rudolph,
  Hajihoseini, Raadu, Gudmundsson, Brenning, Minea, Anders, and
  Lundin}}]{rudolph21:033303}
\bibinfo{author}{\bibfnamefont{M.}~\bibnamefont{Rudolph}},
  \bibinfo{author}{\bibfnamefont{H.}~\bibnamefont{Hajihoseini}},
  \bibinfo{author}{\bibfnamefont{M.~A.} \bibnamefont{Raadu}},
  \bibinfo{author}{\bibfnamefont{J.~T.} \bibnamefont{Gudmundsson}},
  \bibinfo{author}{\bibfnamefont{N.}~\bibnamefont{Brenning}},
  \bibinfo{author}{\bibfnamefont{T.~M.} \bibnamefont{Minea}},
  \bibinfo{author}{\bibfnamefont{A.}~\bibnamefont{Anders}}, \bibnamefont{and}
  \bibinfo{author}{\bibfnamefont{D.}~\bibnamefont{Lundin}},
  \bibinfo{journal}{Journal of Applied Physics} \textbf{\bibinfo{volume}{129}},
  \bibinfo{pages}{033303} (\bibinfo{year}{2021}).

\bibitem[{\citenamefont{Anders}(2008)}]{anders08:201501}
\bibinfo{author}{\bibfnamefont{A.}~\bibnamefont{Anders}},
  \bibinfo{journal}{Applied Physics Letters} \textbf{\bibinfo{volume}{92}},
  \bibinfo{pages}{201501} (\bibinfo{year}{2008}).

\bibitem[{\citenamefont{Gallian et~al.}(2015)\citenamefont{Gallian,
  Trieschmann, Mussenbrock, Brinkmann, and Hitchon}}]{gallian15:023305}
\bibinfo{author}{\bibfnamefont{S.}~\bibnamefont{Gallian}},
  \bibinfo{author}{\bibfnamefont{J.}~\bibnamefont{Trieschmann}},
  \bibinfo{author}{\bibfnamefont{T.}~\bibnamefont{Mussenbrock}},
  \bibinfo{author}{\bibfnamefont{R.~P.} \bibnamefont{Brinkmann}},
  \bibnamefont{and} \bibinfo{author}{\bibfnamefont{W.~N.~G.}
  \bibnamefont{Hitchon}}, \bibinfo{journal}{Journal of Applied Physics}
  \textbf{\bibinfo{volume}{117}}, \bibinfo{pages}{023305}
  (\bibinfo{year}{2015}).

\bibitem[{\citenamefont{Revel et~al.}(2018)\citenamefont{Revel, Minea, and
  Costin}}]{revel18:105009}
\bibinfo{author}{\bibfnamefont{A.}~\bibnamefont{Revel}},
  \bibinfo{author}{\bibfnamefont{T.}~\bibnamefont{Minea}}, \bibnamefont{and}
  \bibinfo{author}{\bibfnamefont{C.}~\bibnamefont{Costin}},
  \bibinfo{journal}{Plasma Sources Science and Technology}
  \textbf{\bibinfo{volume}{27}}, \bibinfo{pages}{105009}
  (\bibinfo{year}{2018}).

\bibitem[{\citenamefont{Revel et~al.}(2016)\citenamefont{Revel, Minea, and
  Tsikata}}]{revel16:100701}
\bibinfo{author}{\bibfnamefont{A.}~\bibnamefont{Revel}},
  \bibinfo{author}{\bibfnamefont{T.}~\bibnamefont{Minea}}, \bibnamefont{and}
  \bibinfo{author}{\bibfnamefont{S.}~\bibnamefont{Tsikata}},
  \bibinfo{journal}{Physics of Plasmas} \textbf{\bibinfo{volume}{23}},
  \bibinfo{pages}{100701} (\bibinfo{year}{2016}).

\bibitem[{\citenamefont{Guimar{\~a}es and Bretagne}(1993)}]{guimaraes93:127}
\bibinfo{author}{\bibfnamefont{F.}~\bibnamefont{Guimar{\~a}es}}
  \bibnamefont{and} \bibinfo{author}{\bibfnamefont{J.}~\bibnamefont{Bretagne}},
  \bibinfo{journal}{Plasma Sources Science and Technology}
  \textbf{\bibinfo{volume}{2}}, \bibinfo{pages}{127} (\bibinfo{year}{1993}).

\bibitem[{\citenamefont{Bretagne et~al.}(1981)\citenamefont{Bretagne, Delouya,
  and Puech}}]{bretagne81:1225}
\bibinfo{author}{\bibfnamefont{J.}~\bibnamefont{Bretagne}},
  \bibinfo{author}{\bibfnamefont{G.}~\bibnamefont{Delouya}}, \bibnamefont{and}
  \bibinfo{author}{\bibfnamefont{V.}~\bibnamefont{Puech}},
  \bibinfo{journal}{Journal of Physics D: Applied Physics}
  \textbf{\bibinfo{volume}{14}}, \bibinfo{pages}{1225} (\bibinfo{year}{1981}).

\bibitem[{\citenamefont{Bretagne et~al.}(1982)\citenamefont{Bretagne, Godart,
  and Puech}}]{bretagne82:2205}
\bibinfo{author}{\bibfnamefont{J.}~\bibnamefont{Bretagne}},
  \bibinfo{author}{\bibfnamefont{J.}~\bibnamefont{Godart}}, \bibnamefont{and}
  \bibinfo{author}{\bibfnamefont{V.}~\bibnamefont{Puech}},
  \bibinfo{journal}{Journal of Physics D: Applied Physics}
  \textbf{\bibinfo{volume}{15}}, \bibinfo{pages}{2205} (\bibinfo{year}{1982}).

\bibitem[{\citenamefont{Bretagne
  et~al.}(1986{\natexlab{a}})\citenamefont{Bretagne, Callede, Legentil, and
  Puech}}]{bretagne86:761}
\bibinfo{author}{\bibfnamefont{J.}~\bibnamefont{Bretagne}},
  \bibinfo{author}{\bibfnamefont{G.}~\bibnamefont{Callede}},
  \bibinfo{author}{\bibfnamefont{M.}~\bibnamefont{Legentil}}, \bibnamefont{and}
  \bibinfo{author}{\bibfnamefont{V.}~\bibnamefont{Puech}},
  \bibinfo{journal}{Journal of Physics D: Applied Physics}
  \textbf{\bibinfo{volume}{19}}, \bibinfo{pages}{761}
  (\bibinfo{year}{1986}{\natexlab{a}}).

\bibitem[{\citenamefont{Guimar{\~a}es et~al.}(1991)\citenamefont{Guimar{\~a}es,
  Almeida, and Bretagne}}]{guimaraes91:133}
\bibinfo{author}{\bibfnamefont{F.}~\bibnamefont{Guimar{\~a}es}},
  \bibinfo{author}{\bibfnamefont{J.}~\bibnamefont{Almeida}}, \bibnamefont{and}
  \bibinfo{author}{\bibfnamefont{J.}~\bibnamefont{Bretagne}},
  \bibinfo{journal}{Journal of Vacuum Science and Technology A}
  \textbf{\bibinfo{volume}{9}}, \bibinfo{pages}{133} (\bibinfo{year}{1991}).

\bibitem[{\citenamefont{Trennepohl et~al.}(1996)\citenamefont{Trennepohl,
  Bretagne, Gousset, Pagnon, and Touzeau}}]{trennepohl96:607}
\bibinfo{author}{\bibfnamefont{W.}~\bibnamefont{Trennepohl}},
  \bibinfo{author}{\bibfnamefont{J.}~\bibnamefont{Bretagne}},
  \bibinfo{author}{\bibfnamefont{G.}~\bibnamefont{Gousset}},
  \bibinfo{author}{\bibfnamefont{D.}~\bibnamefont{Pagnon}}, \bibnamefont{and}
  \bibinfo{author}{\bibfnamefont{M.}~\bibnamefont{Touzeau}},
  \bibinfo{journal}{Plasma Sources Science and Technology}
  \textbf{\bibinfo{volume}{5}}, \bibinfo{pages}{607} (\bibinfo{year}{1996}).

\bibitem[{\citenamefont{Bretagne et~al.}(2015)\citenamefont{Bretagne, Vitelaru,
  Fromy, and Minea}}]{bretagne15:P1.17}
\bibinfo{author}{\bibfnamefont{J.}~\bibnamefont{Bretagne}},
  \bibinfo{author}{\bibfnamefont{C.}~\bibnamefont{Vitelaru}},
  \bibinfo{author}{\bibfnamefont{P.}~\bibnamefont{Fromy}}, \bibnamefont{and}
  \bibinfo{author}{\bibfnamefont{T.}~\bibnamefont{Minea}}, in
  \emph{\bibinfo{booktitle}{Proceedings of the XXXII International Conference
  on Phenomena in Ionized Gases (ICPIG XXXII), Iasi, Romania, July 26 -- 31}}
  (\bibinfo{year}{2015}), p. \bibinfo{pages}{P1.17}.

\bibitem[{\citenamefont{Huo et~al.}(2012)\citenamefont{Huo, Raadu, Lundin,
  Gudmundsson, Anders, and Brenning}}]{huo12:045004}
\bibinfo{author}{\bibfnamefont{C.}~\bibnamefont{Huo}},
  \bibinfo{author}{\bibfnamefont{M.~A.} \bibnamefont{Raadu}},
  \bibinfo{author}{\bibfnamefont{D.}~\bibnamefont{Lundin}},
  \bibinfo{author}{\bibfnamefont{J.~T.} \bibnamefont{Gudmundsson}},
  \bibinfo{author}{\bibfnamefont{A.}~\bibnamefont{Anders}}, \bibnamefont{and}
  \bibinfo{author}{\bibfnamefont{N.}~\bibnamefont{Brenning}},
  \bibinfo{journal}{Plasma Sources Science and Technology}
  \textbf{\bibinfo{volume}{21}}, \bibinfo{pages}{045004}
  (\bibinfo{year}{2012}).

\bibitem[{\citenamefont{Brenning et~al.}(2012)\citenamefont{Brenning, Lundin,
  Raadu, Huo, Vitelaru, Stancu, Minea, and Helmersson}}]{brenning12:025005}
\bibinfo{author}{\bibfnamefont{N.}~\bibnamefont{Brenning}},
  \bibinfo{author}{\bibfnamefont{D.}~\bibnamefont{Lundin}},
  \bibinfo{author}{\bibfnamefont{M.~A.} \bibnamefont{Raadu}},
  \bibinfo{author}{\bibfnamefont{C.}~\bibnamefont{Huo}},
  \bibinfo{author}{\bibfnamefont{C.}~\bibnamefont{Vitelaru}},
  \bibinfo{author}{\bibfnamefont{G.~D.} \bibnamefont{Stancu}},
  \bibinfo{author}{\bibfnamefont{T.}~\bibnamefont{Minea}}, \bibnamefont{and}
  \bibinfo{author}{\bibfnamefont{U.}~\bibnamefont{Helmersson}},
  \bibinfo{journal}{Plasma Sources Science and Technology}
  \textbf{\bibinfo{volume}{21}}, \bibinfo{pages}{025005}
  (\bibinfo{year}{2012}).

\bibitem[{\citenamefont{Stancu et~al.}(2015)\citenamefont{Stancu, Brenning,
  Vitelaru, Lundin, and Minea}}]{stancu15:045011}
\bibinfo{author}{\bibfnamefont{G.~D.} \bibnamefont{Stancu}},
  \bibinfo{author}{\bibfnamefont{N.}~\bibnamefont{Brenning}},
  \bibinfo{author}{\bibfnamefont{C.}~\bibnamefont{Vitelaru}},
  \bibinfo{author}{\bibfnamefont{D.}~\bibnamefont{Lundin}}, \bibnamefont{and}
  \bibinfo{author}{\bibfnamefont{T.}~\bibnamefont{Minea}},
  \bibinfo{journal}{Plasma Sources Science and Technology}
  \textbf{\bibinfo{volume}{24}}, \bibinfo{pages}{045011}
  (\bibinfo{year}{2015}).

\bibitem[{\citenamefont{Gudmundsson et~al.}(2015)\citenamefont{Gudmundsson,
  Lundin, Stancu, Brenning, and Minea}}]{gudmundsson15:113508}
\bibinfo{author}{\bibfnamefont{J.~T.} \bibnamefont{Gudmundsson}},
  \bibinfo{author}{\bibfnamefont{D.}~\bibnamefont{Lundin}},
  \bibinfo{author}{\bibfnamefont{G.~D.} \bibnamefont{Stancu}},
  \bibinfo{author}{\bibfnamefont{N.}~\bibnamefont{Brenning}}, \bibnamefont{and}
  \bibinfo{author}{\bibfnamefont{T.~M.} \bibnamefont{Minea}},
  \bibinfo{journal}{Physics of Plasmas} \textbf{\bibinfo{volume}{22}},
  \bibinfo{pages}{113508} (\bibinfo{year}{2015}).

\bibitem[{\citenamefont{Rudolph et~al.}(2020)\citenamefont{Rudolph, Brenning,
  Raadu, Hajihoseini, Gudmundsson, Anders, and Lundin}}]{rudolph20:05LT01}
\bibinfo{author}{\bibfnamefont{M.}~\bibnamefont{Rudolph}},
  \bibinfo{author}{\bibfnamefont{N.}~\bibnamefont{Brenning}},
  \bibinfo{author}{\bibfnamefont{M.~A.} \bibnamefont{Raadu}},
  \bibinfo{author}{\bibfnamefont{H.}~\bibnamefont{Hajihoseini}},
  \bibinfo{author}{\bibfnamefont{J.~T.} \bibnamefont{Gudmundsson}},
  \bibinfo{author}{\bibfnamefont{A.}~\bibnamefont{Anders}}, \bibnamefont{and}
  \bibinfo{author}{\bibfnamefont{D.}~\bibnamefont{Lundin}},
  \bibinfo{journal}{Plasma Sources Science and Technology}
  \textbf{\bibinfo{volume}{29}}, \bibinfo{pages}{05LT01}
  (\bibinfo{year}{2020}).

\bibitem[{\citenamefont{Butler et~al.}(2018)\citenamefont{Butler, Brenning,
  Raadu, Gudmundsson, Minea, and Lundin}}]{butler18:105005}
\bibinfo{author}{\bibfnamefont{A.}~\bibnamefont{Butler}},
  \bibinfo{author}{\bibfnamefont{N.}~\bibnamefont{Brenning}},
  \bibinfo{author}{\bibfnamefont{M.~A.} \bibnamefont{Raadu}},
  \bibinfo{author}{\bibfnamefont{J.~T.} \bibnamefont{Gudmundsson}},
  \bibinfo{author}{\bibfnamefont{T.}~\bibnamefont{Minea}}, \bibnamefont{and}
  \bibinfo{author}{\bibfnamefont{D.}~\bibnamefont{Lundin}},
  \bibinfo{journal}{Plasma Sources Science and Technology}
  \textbf{\bibinfo{volume}{27}}, \bibinfo{pages}{105005}
  (\bibinfo{year}{2018}).

\bibitem[{\citenamefont{Gudmundsson et~al.}(2016)\citenamefont{Gudmundsson,
  Lundin, Brenning, Raadu, Huo, and Minea}}]{gudmundsson16:065004}
\bibinfo{author}{\bibfnamefont{J.~T.} \bibnamefont{Gudmundsson}},
  \bibinfo{author}{\bibfnamefont{D.}~\bibnamefont{Lundin}},
  \bibinfo{author}{\bibfnamefont{N.}~\bibnamefont{Brenning}},
  \bibinfo{author}{\bibfnamefont{M.~A.} \bibnamefont{Raadu}},
  \bibinfo{author}{\bibfnamefont{C.}~\bibnamefont{Huo}}, \bibnamefont{and}
  \bibinfo{author}{\bibfnamefont{T.~M.} \bibnamefont{Minea}},
  \bibinfo{journal}{Plasma Sources Science and Technology}
  \textbf{\bibinfo{volume}{25}}, \bibinfo{pages}{065004}
  (\bibinfo{year}{2016}).

\bibitem[{IST()}]{ISTLisbon}
\emph{\bibinfo{title}{{IST}-{Lisbon} database, {\tt www.lxcat.net/ist-lisbon},
  \mbox{retrieved on {M}ay 18, 2020}}}.

\bibitem[{\citenamefont{Alves}(2014)}]{alves14:012007}
\bibinfo{author}{\bibfnamefont{L.~L.} \bibnamefont{Alves}},
  \bibinfo{journal}{Journal of Physics: Conference Series}
  \textbf{\bibinfo{volume}{565}}, \bibinfo{pages}{012007}
  (\bibinfo{year}{2014}).

\bibitem[{\citenamefont{Yanguas-Gil et~al.}(2005)\citenamefont{Yanguas-Gil,
  Cotrino, and Alves}}]{yanguasgil05:1588}
\bibinfo{author}{\bibfnamefont{A.}~\bibnamefont{Yanguas-Gil}},
  \bibinfo{author}{\bibfnamefont{J.}~\bibnamefont{Cotrino}}, \bibnamefont{and}
  \bibinfo{author}{\bibfnamefont{L.~L.} \bibnamefont{Alves}},
  \bibinfo{journal}{Journal of Physics D: Applied Physics}
  \textbf{\bibinfo{volume}{38}}, \bibinfo{pages}{1588} (\bibinfo{year}{2005}).

\bibitem[{\citenamefont{Khakoo et~al.}(2004)\citenamefont{Khakoo, Vandeventer,
  Childers, Kanik, Fontes, Bartschat, Zeman, Madison, Saxena, Srivastava
  and Stauffer}}]{khakoo04:247}
\bibinfo{author}{\bibfnamefont{M.~A.} \bibnamefont{Khakoo}},
  \bibinfo{author}{\bibfnamefont{P.}~\bibnamefont{Vandeventer}},
  \bibinfo{author}{\bibfnamefont{J.~G.} \bibnamefont{Childers}},
  \bibinfo{author}{\bibfnamefont{I.}~\bibnamefont{Kanik}},
  \bibinfo{author}{\bibfnamefont{C.~J.} \bibnamefont{Fontes}},
  \bibinfo{author}{\bibfnamefont{K.}~\bibnamefont{Bartschat}},
  \bibinfo{author}{\bibfnamefont{V.}~\bibnamefont{Zeman}},
  \bibinfo{author}{\bibfnamefont{D.~H.} \bibnamefont{Madison}},
  \bibinfo{author}{\bibfnamefont{S.}~\bibnamefont{Saxena}},
  \bibinfo{author}{\bibfnamefont{R.}~\bibnamefont{Srivastava}}, \bibnamefont{and}
  \bibinfo{author}{\bibfnamefont{A.~D.}~\bibnamefont{Stauffer}}, 
  \bibinfo{journal}{Journal of Physics B: Atomic,
  Molecular and Optical Physics} \textbf{\bibinfo{volume}{37}},
  \bibinfo{pages}{247} (\bibinfo{year}{2004}).

\bibitem[{\citenamefont{Dixon et~al.}(1973)\citenamefont{Dixon, Harrison, and
  Smith}}]{dixon73:405}
\bibinfo{author}{\bibfnamefont{A.~J.} \bibnamefont{Dixon}},
  \bibinfo{author}{\bibfnamefont{M.~F.~A.} \bibnamefont{Harrison}},
  \bibnamefont{and} \bibinfo{author}{\bibfnamefont{A.~C.~H.}
  \bibnamefont{Smith}}, in \emph{\bibinfo{booktitle}{8th International
  Conference on the Physics of Electronic and Atomic Collisions (VIII
  ICPEAC)}}, edited by \bibinfo{editor}{\bibfnamefont{B.~C.}
  \bibnamefont{Cobi{\'c}}} \bibnamefont{and}
  \bibinfo{editor}{\bibfnamefont{M.~V.} \bibnamefont{Kurepa}}
  (\bibinfo{publisher}{Institute of Physics}, \bibinfo{address}{Belgrade,
  Yugoslavia}, \bibinfo{year}{1973}), pp. \bibinfo{pages}{405--406}.

\bibitem[{\citenamefont{Freund}(1987)}]{freund87:329}
\bibinfo{author}{\bibfnamefont{R.~S.} \bibnamefont{Freund}}, in
  \emph{\bibinfo{booktitle}{Swarm Studies and Inelastic Electron-Molecule
  Collisions Proceedings of the Meeting of the Fourth International Swarm
  Seminar and the Inelastic Electron-Molecule Collisions Symposium, July
  19--23, 1985, Tahoe City, California, USA}}, edited by
  \bibinfo{editor}{\bibfnamefont{L.~C.} \bibnamefont{Pitchford}},
  \bibinfo{editor}{\bibfnamefont{B.~V.} \bibnamefont{{McKoy}}},
  \bibinfo{editor}{\bibfnamefont{A.}~\bibnamefont{Chutjian}}, \bibnamefont{and}
  \bibinfo{editor}{\bibfnamefont{S.}~\bibnamefont{Trajrnar}}
  (\bibinfo{publisher}{Springer-Verlag}, \bibinfo{address}{New York},
  \bibinfo{year}{1987}), pp. \bibinfo{pages}{329--346}.

\bibitem[{\citenamefont{Lieberman and Lichtenberg}(2005)}]{lieberman05}
\bibinfo{author}{\bibfnamefont{M.~A.} \bibnamefont{Lieberman}}
  \bibnamefont{and} \bibinfo{author}{\bibfnamefont{A.~J.}
  \bibnamefont{Lichtenberg}}, \emph{\bibinfo{title}{Principles of \mbox{P}lasma
  \mbox{D}ischarges and \mbox{M}aterials \mbox{P}rocessing}}
  (\bibinfo{publisher}{John Wiley $\&$ Sons}, \bibinfo{address}{New York},
  \bibinfo{year}{2005}), \bibinfo{edition}{2nd} ed.

\bibitem[{\citenamefont{Gudmundsson and
  Thorsteinsson}(2007)}]{gudmundsson07:399}
\bibinfo{author}{\bibfnamefont{J.~T.} \bibnamefont{Gudmundsson}}
  \bibnamefont{and} \bibinfo{author}{\bibfnamefont{E.~G.}
  \bibnamefont{Thorsteinsson}}, \bibinfo{journal}{Plasma Sources Science and
  Technology} \textbf{\bibinfo{volume}{16}}, \bibinfo{pages}{399}
  (\bibinfo{year}{2007}).

\bibitem[{\citenamefont{Deutsch et~al.}(2008)\citenamefont{Deutsch, Becker, and
  M\"{a}rk}}]{deutsch08:58}
\bibinfo{author}{\bibfnamefont{H.}~\bibnamefont{Deutsch}},
  \bibinfo{author}{\bibfnamefont{K.}~\bibnamefont{Becker}}, \bibnamefont{and}
  \bibinfo{author}{\bibfnamefont{T.}~\bibnamefont{M\"{a}rk}},
  \bibinfo{journal}{International Journal of Mass Spectrometry}
  \textbf{\bibinfo{volume}{271}}, \bibinfo{pages}{58 } (\bibinfo{year}{2008}).

\bibitem[{\citenamefont{Bartlett and Stelbovics}(2004)}]{bartlett04:235}
\bibinfo{author}{\bibfnamefont{P.~L.} \bibnamefont{Bartlett}} \bibnamefont{and}
  \bibinfo{author}{\bibfnamefont{A.~T.} \bibnamefont{Stelbovics}},
  \bibinfo{journal}{Atomic Data and Nuclear Data Tables}
  \textbf{\bibinfo{volume}{86}}, \bibinfo{pages}{235 } (\bibinfo{year}{2004}).

\bibitem[{\citenamefont{Diserens et~al.}(1988)\citenamefont{Diserens, Smith,
  and Harrison}}]{diserens88:2129}
\bibinfo{author}{\bibfnamefont{M.~J.} \bibnamefont{Diserens}},
  \bibinfo{author}{\bibfnamefont{A.~C.~H.} \bibnamefont{Smith}},
  \bibnamefont{and} \bibinfo{author}{\bibfnamefont{M.~F.~A.}
  \bibnamefont{Harrison}}, \bibinfo{journal}{Journal of Physics B: Atomic,
  Molecular and Optical Physics} \textbf{\bibinfo{volume}{21}},
  \bibinfo{pages}{2129} (\bibinfo{year}{1988}).

\bibitem[{\citenamefont{Riseberg et~al.}(1973)\citenamefont{Riseberg, Parks,
  and Schearer}}]{riseberg73:1962}
\bibinfo{author}{\bibfnamefont{L.~A.} \bibnamefont{Riseberg}},
  \bibinfo{author}{\bibfnamefont{W.~F.} \bibnamefont{Parks}}, \bibnamefont{and}
  \bibinfo{author}{\bibfnamefont{L.~D.} \bibnamefont{Schearer}},
  \bibinfo{journal}{Physical Review A} \textbf{\bibinfo{volume}{8}},
  \bibinfo{pages}{1962 } (\bibinfo{year}{1973}).

\bibitem[{\citenamefont{Bretagne
  et~al.}(1986{\natexlab{b}})\citenamefont{Bretagne, Callede, Legentil, and
  Puech}}]{bretagne86:779}
\bibinfo{author}{\bibfnamefont{J.}~\bibnamefont{Bretagne}},
  \bibinfo{author}{\bibfnamefont{G.}~\bibnamefont{Callede}},
  \bibinfo{author}{\bibfnamefont{M.}~\bibnamefont{Legentil}}, \bibnamefont{and}
  \bibinfo{author}{\bibfnamefont{V.}~\bibnamefont{Puech}},
  \bibinfo{journal}{Journal of Physics D: Applied Physics}
  \textbf{\bibinfo{volume}{19}}, \bibinfo{pages}{779}
  (\bibinfo{year}{1986}{\natexlab{b}}).

\bibitem[{\citenamefont{Vl\v{c}ek}(1989)}]{vlcek89:623}
\bibinfo{author}{\bibfnamefont{J.}~\bibnamefont{Vl\v{c}ek}},
  \bibinfo{journal}{Journal of Physics D: Applied Physics}
  \textbf{\bibinfo{volume}{22}}, \bibinfo{pages}{623} (\bibinfo{year}{1989}).

\bibitem[{\citenamefont{Rockwood}(1973)}]{rockwood73:2348}
\bibinfo{author}{\bibfnamefont{S.~D.} \bibnamefont{Rockwood}},
  \bibinfo{journal}{Physical Review A} \textbf{\bibinfo{volume}{8}},
  \bibinfo{pages}{2348} (\bibinfo{year}{1973}).

\bibitem[{\citenamefont{Vl\v{c}ek and Pelik{\'a}n}(1989)}]{vlcek89:632}
\bibinfo{author}{\bibfnamefont{J.}~\bibnamefont{Vl\v{c}ek}} \bibnamefont{and}
  \bibinfo{author}{\bibfnamefont{V.}~\bibnamefont{Pelik{\'a}n}},
  \bibinfo{journal}{Journal of Physics D: Applied Physics}
  \textbf{\bibinfo{volume}{22}}, \bibinfo{pages}{632} (\bibinfo{year}{1989}).

\bibitem[{\citenamefont{Vl\v{c}ek and Pelik{\'a}n}(1990)}]{vlcek90:526}
\bibinfo{author}{\bibfnamefont{J.}~\bibnamefont{Vl\v{c}ek}} \bibnamefont{and}
  \bibinfo{author}{\bibfnamefont{V.}~\bibnamefont{Pelik{\'a}n}},
  \bibinfo{journal}{Journal of Physics D: Applied Physics}
  \textbf{\bibinfo{volume}{23}}, \bibinfo{pages}{526} (\bibinfo{year}{1990}).

\bibitem[{\citenamefont{Bogaerts et~al.}(1998)\citenamefont{Bogaerts, Gijbels,
  and Vl\v{c}ek}}]{bogaerts98:121}
\bibinfo{author}{\bibfnamefont{A.}~\bibnamefont{Bogaerts}},
  \bibinfo{author}{\bibfnamefont{R.}~\bibnamefont{Gijbels}}, \bibnamefont{and}
  \bibinfo{author}{\bibfnamefont{J.}~\bibnamefont{Vl\v{c}ek}},
  \bibinfo{journal}{Journal of Applied Physics} \textbf{\bibinfo{volume}{84}},
  \bibinfo{pages}{121} (\bibinfo{year}{1998}).

\bibitem[{\citenamefont{Bultel et~al.}(2002)\citenamefont{Bultel, van Ootegem,
  Bourdon, and Vervisch}}]{bultel02:046406}
\bibinfo{author}{\bibfnamefont{A.}~\bibnamefont{Bultel}},
  \bibinfo{author}{\bibfnamefont{B.}~\bibnamefont{van Ootegem}},
  \bibinfo{author}{\bibfnamefont{A.}~\bibnamefont{Bourdon}}, \bibnamefont{and}
  \bibinfo{author}{\bibfnamefont{P.}~\bibnamefont{Vervisch}},
  \bibinfo{journal}{Physical Review E} \textbf{\bibinfo{volume}{65}},
  \bibinfo{pages}{046406} (\bibinfo{year}{2002}).

\bibitem[{\citenamefont{Akatsuka}(2009)}]{akatsuka09:043502}
\bibinfo{author}{\bibfnamefont{H.}~\bibnamefont{Akatsuka}},
  \bibinfo{journal}{Physics of Plasmas} \textbf{\bibinfo{volume}{16}},
  \bibinfo{pages}{043502} (\bibinfo{year}{2009}).

\bibitem[{\citenamefont{Drawin and Katsonis}(1976)}]{drawin76rr}
\bibinfo{author}{\bibfnamefont{H.~W.} \bibnamefont{Drawin}} \bibnamefont{and}
  \bibinfo{author}{\bibfnamefont{K.}~\bibnamefont{Katsonis}},
  \bibinfo{type}{Tech. Rep.}, \bibinfo{institution}{European Atomic Energy
  Community. Commissariat a l'Energie Atomique, EUR-CEA-PC-837},
  \bibinfo{address}{Fontenay-aux-Roses, France} (\bibinfo{year}{1976}).

\bibitem[{\citenamefont{Drawin}(1967)}]{drawin67rr}
\bibinfo{author}{\bibfnamefont{H.~W.} \bibnamefont{Drawin}},
  \bibinfo{type}{Tech. Rep.}, \bibinfo{institution}{European Atomic Energy
  Community. Commissariat a l'Energie Atomique, EUR-CEA-FC-383},
  \bibinfo{address}{Fontenay-aux-Roses, France} (\bibinfo{year}{1967}).

\bibitem[{\citenamefont{Bethe}(1932)}]{bethe32:293}
\bibinfo{author}{\bibfnamefont{H.}~\bibnamefont{Bethe}},
  \bibinfo{journal}{Zeitschrift f{\"u}r Physik} \textbf{\bibinfo{volume}{76}},
  \bibinfo{pages}{293} (\bibinfo{year}{1932}).

\bibitem[{\citenamefont{Inokuti}(1971)}]{inokuti71:297}
\bibinfo{author}{\bibfnamefont{M.}~\bibnamefont{Inokuti}},
  \bibinfo{journal}{Reviews of Modern Physics} \textbf{\bibinfo{volume}{43}},
  \bibinfo{pages}{297} (\bibinfo{year}{1971}).

\bibitem[{\citenamefont{Drawin}(1969)}]{drawin69:483}
\bibinfo{author}{\bibfnamefont{H.~W.} \bibnamefont{Drawin}},
  \bibinfo{journal}{Zeitschrift f{\"u}r Physik} \textbf{\bibinfo{volume}{225}},
  \bibinfo{pages}{483} (\bibinfo{year}{1969}).

\bibitem[{\citenamefont{Drawin}(1961)}]{drawin61:513}
\bibinfo{author}{\bibfnamefont{H.-W.} \bibnamefont{Drawin}},
  \bibinfo{journal}{Zeitschrift f{\"u}r Physik} \textbf{\bibinfo{volume}{164}}
  (\bibinfo{year}{1961}).

\bibitem[{\citenamefont{Straub et~al.}(1995)\citenamefont{Straub, Renault,
  Lindsay, Smith, and Stebbings}}]{straub95:1115}
\bibinfo{author}{\bibfnamefont{H.~C.} \bibnamefont{Straub}},
  \bibinfo{author}{\bibfnamefont{P.}~\bibnamefont{Renault}},
  \bibinfo{author}{\bibfnamefont{B.~G.} \bibnamefont{Lindsay}},
  \bibinfo{author}{\bibfnamefont{K.~A.} \bibnamefont{Smith}}, \bibnamefont{and}
  \bibinfo{author}{\bibfnamefont{R.~F.} \bibnamefont{Stebbings}},
  \bibinfo{journal}{Physical Review A} \textbf{\bibinfo{volume}{52}},
  \bibinfo{pages}{1115 } (\bibinfo{year}{1995}).

\bibitem[{\citenamefont{Frost and Phelps}(1964)}]{frost64:A1538}
\bibinfo{author}{\bibfnamefont{L.~S.} \bibnamefont{Frost}} \bibnamefont{and}
  \bibinfo{author}{\bibfnamefont{A.~V.} \bibnamefont{Phelps}},
  \bibinfo{journal}{Physical Review} \textbf{\bibinfo{volume}{136}},
  \bibinfo{pages}{A1538} (\bibinfo{year}{1964}).

\bibitem[{\citenamefont{Huo et~al.}(2014)\citenamefont{Huo, Lundin, Raadu,
  Anders, Gudmundsson, and Brenning}}]{huo14:025017}
\bibinfo{author}{\bibfnamefont{C.}~\bibnamefont{Huo}},
  \bibinfo{author}{\bibfnamefont{D.}~\bibnamefont{Lundin}},
  \bibinfo{author}{\bibfnamefont{M.~A.} \bibnamefont{Raadu}},
  \bibinfo{author}{\bibfnamefont{A.}~\bibnamefont{Anders}},
  \bibinfo{author}{\bibfnamefont{J.~T.} \bibnamefont{Gudmundsson}},
  \bibnamefont{and} \bibinfo{author}{\bibfnamefont{N.}~\bibnamefont{Brenning}},
  \bibinfo{journal}{Plasma Sources Science and Technology}
  \textbf{\bibinfo{volume}{23}}, \bibinfo{pages}{025017}
  (\bibinfo{year}{2014}).

\bibitem[{\citenamefont{Hajihoseini et~al.}(2019)\citenamefont{Hajihoseini,
  \v{C}ada, Hubi\v{c}ka, {\"U}naldi, Raadu, Brenning, Gudmundsson, and
  Lundin}}]{hajihoseini19:201}
\bibinfo{author}{\bibfnamefont{H.}~\bibnamefont{Hajihoseini}},
  \bibinfo{author}{\bibfnamefont{M.}~\bibnamefont{\v{C}ada}},
  \bibinfo{author}{\bibfnamefont{Z.}~\bibnamefont{Hubi\v{c}ka}},
  \bibinfo{author}{\bibfnamefont{S.}~\bibnamefont{{\"U}naldi}},
  \bibinfo{author}{\bibfnamefont{M.~A.} \bibnamefont{Raadu}},
  \bibinfo{author}{\bibfnamefont{N.}~\bibnamefont{Brenning}},
  \bibinfo{author}{\bibfnamefont{J.~T.} \bibnamefont{Gudmundsson}},
  \bibnamefont{and} \bibinfo{author}{\bibfnamefont{D.}~\bibnamefont{Lundin}},
  \bibinfo{journal}{Plasma} \textbf{\bibinfo{volume}{2}}, \bibinfo{pages}{201}
  (\bibinfo{year}{2019}).

\bibitem[{\citenamefont{Hajihoseini et~al.}(2020)\citenamefont{Hajihoseini,
  \v{C}ada, Hubi\v{c}ka, {\"U}naldi, Raadu, Brenning, Gudmundsson, and
  Lundin}}]{hajihoseini20:033009}
\bibinfo{author}{\bibfnamefont{H.}~\bibnamefont{Hajihoseini}},
  \bibinfo{author}{\bibfnamefont{M.}~\bibnamefont{\v{C}ada}},
  \bibinfo{author}{\bibfnamefont{Z.}~\bibnamefont{Hubi\v{c}ka}},
  \bibinfo{author}{\bibfnamefont{S.}~\bibnamefont{{\"U}naldi}},
  \bibinfo{author}{\bibfnamefont{M.~A.} \bibnamefont{Raadu}},
  \bibinfo{author}{\bibfnamefont{N.}~\bibnamefont{Brenning}},
  \bibinfo{author}{\bibfnamefont{J.~T.} \bibnamefont{Gudmundsson}},
  \bibnamefont{and} \bibinfo{author}{\bibfnamefont{D.}~\bibnamefont{Lundin}},
  \bibinfo{journal}{Journal of Vacuum Science and Technology A}
  \textbf{\bibinfo{volume}{38}}, \bibinfo{pages}{033009}
  (\bibinfo{year}{2020}).

\bibitem[{\citenamefont{Kubart et~al.}(2014)\citenamefont{Kubart, \v{C}ada,
  Lundin, and Hubi\v{c}ka}}]{kubart14:152}
\bibinfo{author}{\bibfnamefont{T.}~\bibnamefont{Kubart}},
  \bibinfo{author}{\bibfnamefont{M.}~\bibnamefont{\v{C}ada}},
  \bibinfo{author}{\bibfnamefont{D.}~\bibnamefont{Lundin}}, \bibnamefont{and}
  \bibinfo{author}{\bibfnamefont{Z.}~\bibnamefont{Hubi\v{c}ka}},
  \bibinfo{journal}{Surface and Coatings Technology}
  \textbf{\bibinfo{volume}{238}}, \bibinfo{pages}{152} (\bibinfo{year}{2014}).

\bibitem[{\citenamefont{Katsonis et~al.}(2011)\citenamefont{Katsonis,
  Berenguer, Kaminska, and Dudeck}}]{katsonis11:896836}
\bibinfo{author}{\bibfnamefont{K.}~\bibnamefont{Katsonis}},
  \bibinfo{author}{\bibfnamefont{C.}~\bibnamefont{Berenguer}},
  \bibinfo{author}{\bibfnamefont{A.}~\bibnamefont{Kaminska}}, \bibnamefont{and}
  \bibinfo{author}{\bibfnamefont{M.}~\bibnamefont{Dudeck}},
  \bibinfo{journal}{International Journal of Aerospace Engineering}
  \textbf{\bibinfo{volume}{2011}}, \bibinfo{pages}{896836}
  (\bibinfo{year}{2011}).

\bibitem[{\citenamefont{Sigurjonsson et~al.}(2009)\citenamefont{Sigurjonsson,
  Larsson, Lundin, Helmersson, and Gudmundsson}}]{sigurjonsson09:234}
\bibinfo{author}{\bibfnamefont{P.}~\bibnamefont{Sigurjonsson}},
  \bibinfo{author}{\bibfnamefont{P.}~\bibnamefont{Larsson}},
  \bibinfo{author}{\bibfnamefont{D.}~\bibnamefont{Lundin}},
  \bibinfo{author}{\bibfnamefont{U.}~\bibnamefont{Helmersson}},
  \bibnamefont{and} \bibinfo{author}{\bibfnamefont{J.~T.}
  \bibnamefont{Gudmundsson}}, in \emph{\bibinfo{booktitle}{Proceedings of the
  52nd Society of Vacuum Coaters Annual Technical Conferenece, May 9--14, 2009,
  Santa Clara, California}} (\bibinfo{publisher}{Society of Vacuum Coaters},
  \bibinfo{address}{Albuquerque, New Mexico}, \bibinfo{year}{2009}), pp.
  \bibinfo{pages}{234--239}.

\bibitem[{\citenamefont{Gudmundsson et~al.}(2009)\citenamefont{Gudmundsson,
  Sigurjonsson, Larsson, Lundin, and Helmersson}}]{gudmundsson09:123302}
\bibinfo{author}{\bibfnamefont{J.~T.} \bibnamefont{Gudmundsson}},
  \bibinfo{author}{\bibfnamefont{P.}~\bibnamefont{Sigurjonsson}},
  \bibinfo{author}{\bibfnamefont{P.}~\bibnamefont{Larsson}},
  \bibinfo{author}{\bibfnamefont{D.}~\bibnamefont{Lundin}}, \bibnamefont{and}
  \bibinfo{author}{\bibfnamefont{U.}~\bibnamefont{Helmersson}},
  \bibinfo{journal}{Journal of Applied Physics} \textbf{\bibinfo{volume}{105}},
  \bibinfo{pages}{123302} (\bibinfo{year}{2009}).

\bibitem[{\citenamefont{Brenning et~al.}(2020)\citenamefont{Brenning, Butler,
  Hajihoseini, Rudolph, Raadu, Gudmundsson, Minea, and
  Lundin}}]{brenning20:033008}
\bibinfo{author}{\bibfnamefont{N.}~\bibnamefont{Brenning}},
  \bibinfo{author}{\bibfnamefont{A.}~\bibnamefont{Butler}},
  \bibinfo{author}{\bibfnamefont{H.}~\bibnamefont{Hajihoseini}},
  \bibinfo{author}{\bibfnamefont{M.}~\bibnamefont{Rudolph}},
  \bibinfo{author}{\bibfnamefont{M.~A.} \bibnamefont{Raadu}},
  \bibinfo{author}{\bibfnamefont{J.~T.} \bibnamefont{Gudmundsson}},
  \bibinfo{author}{\bibfnamefont{T.}~\bibnamefont{Minea}}, \bibnamefont{and}
  \bibinfo{author}{\bibfnamefont{D.}~\bibnamefont{Lundin}},
  \bibinfo{journal}{Journal of Vacuum Science and Technology A}
  \textbf{\bibinfo{volume}{38}}, \bibinfo{pages}{033008}
  (\bibinfo{year}{2020}).

\bibitem[{\citenamefont{Brenning et~al.}(2021)\citenamefont{Brenning,
  Hajihoseini, Rudolph, Raadu, Gudmundsson, Minea, and
  Lundin}}]{brenning21:015015}
\bibinfo{author}{\bibfnamefont{N.}~\bibnamefont{Brenning}},
  \bibinfo{author}{\bibfnamefont{H.}~\bibnamefont{Hajihoseini}},
  \bibinfo{author}{\bibfnamefont{M.}~\bibnamefont{Rudolph}},
  \bibinfo{author}{\bibfnamefont{M.~A.} \bibnamefont{Raadu}},
  \bibinfo{author}{\bibfnamefont{J.~T.} \bibnamefont{Gudmundsson}},
  \bibinfo{author}{\bibfnamefont{T.~M.} \bibnamefont{Minea}}, \bibnamefont{and}
  \bibinfo{author}{\bibfnamefont{D.}~\bibnamefont{Lundin}},
  \bibinfo{journal}{Plasma Sources Science and Technology}
  \textbf{\bibinfo{volume}{30}}, \bibinfo{pages}{015015}
  (\bibinfo{year}{2021}).

\end{thebibliography}
\end{document}